\documentclass[final,onefignum,onetabnum]{siamart190516}
\usepackage{amsmath}
\usepackage{amssymb}
\usepackage{bm}
\usepackage{bbm}
\usepackage{graphicx}
\usepackage{url}
\usepackage{xcolor}
\usepackage{cite}
\usepackage{enumitem}

\newtheorem{remark}[theorem]{Remark}

\newcommand{\new}[1]{{{\color{black}#1}}}

\title{Tunable Eigenvector-Based Centralities for Multiplex and Temporal Networks
\thanks{We thank Daryl DeFord, Tina Eliassi-Rad, Des Higham, Christine Klymko, Marianne McKenzie, Scott Pauls, Kaiyan Peng, and Michael Schaub for helpful comments. 
DT was supported by the Simons Foundation under Award \#578333. 
MAP was supported by the National Science Foundation (grant \#1922952) through the Algorithms for Threat Detection (ATD) program.
PJM was supported by the Eunice Kennedy Shriver National Institute of Child Health \& Human Development of the National Institutes of Health under Award Number R01HD075712 and by the James S. McDonnell Foundation 21st Century Science Initiative - Complex Systems Scholar Award grant \#220020315. 
The content is solely the responsibility of the authors and does not necessarily reflect the views of any of the funding agencies.}}

\author{Dane Taylor\thanks{Department of Mathematics, University at Buffalo, State University of New York, Buffalo, NY 14260, USA}
\and Mason A. Porter\thanks{Department of Mathematics, University of California, Los Angeles, CA 90095, USA}
\and Peter J. Mucha\thanks{Carolina Center for Interdisciplinary Applied Mathematics, Department of Mathematics and Department of Applied Physical Sciences, University of North Carolina, Chapel Hill, NC 27599, USA}
}

\begin{document}

\maketitle 

\begin{abstract}
Characterizing the importances (i.e., centralities) of nodes in social, biological, and technological networks is a core topic in both network analysis and data science. We present a linear-algebraic framework that generalizes eigenvector-based centralities, including PageRank and hub/authority scores, to provide a common framework for two popular classes of multilayer networks: multiplex networks (which have layers that encode different types of relationships) and temporal networks (in which the relationships change over time). Our approach involves the study of joint, marginal, and conditional ``supracentralities'' that one can calculate from the dominant eigenvector of a \emph{supracentrality matrix} [Taylor et al., 2017], which couples centrality matrices that are associated with individual network layers. We extend this prior work (which was restricted to temporal networks with layers that are coupled by adjacent-in-time coupling) by allowing the layers to be coupled through a (possibly asymmetric) interlayer-adjacency matrix $\tilde{\bm{A}}$, where the entry $\tilde{A}_{tt'} \geq 0$ encodes the coupling between layers $t$ and $t'$.  Our framework provides a unifying foundation for centrality analysis of multiplex and temporal networks, and it also illustrates a complicated dependency of the supracentralities on the topology and weights of interlayer coupling. By scaling $\tilde{\bm{A}}$ by an interlayer-coupling strength $\omega\ge0$ and developing a singular perturbation theory for the limits of weak  ($\omega\to0^+$) and strong ($\omega\to\infty$) coupling, we also reveal an interesting dependence of supracentralities on the  right and left dominant eigenvectors of $\tilde{\bm{A}}$. We provide additional theoretical and practical insights by applying our framework to two empirical data sets: a multiplex network of airline transportation in Europe and a temporal network that encodes the graduation and hiring of mathematical scientists at United States universities.
\end{abstract}



\pagestyle{myheadings}
\thispagestyle{plain}
\markboth{D. TAYLOR {\it et al.}}{Centrality for  Multiplex and Temporal  Networks}

%
%

\section{Introduction}\label{sec:intro}

Quantifying the importance of entities in a network is an essential feature of many search engines on the World Wide Web \cite{Brin1998conf,pagerank,langville2006,gleich2014}, ranking algorithms for sports teams and athletes \cite{monthly,saavedra2010,chartier2013}, targeted social-network marketing schemes \cite{kempe2003}, investigations of fragility in infrastructures \cite{holme2003congestion,guimera2005worldwide}, quantitative analysis of the impact of research papers and scientists \cite{Fortunato2017}, examinations of the influence of judicial and legislative documents \cite{leicht-citation2007,fowler2007b}, identification of novel drug targets in biological systems \cite{jeong2001lethality}, and many other applications. In the most common (and simplest) type of network, called a ``graph'' or a ``monolayer network'', a node represents an entity (e.g., a web page, a person, a document, or a protein), and an edge encodes a relationship between a pair of entities. \emph{Centrality analysis}, in which one seeks to quantify the importances of nodes and/or edges (and, more generally, of other subgraphs as well), has been developed intensively across numerous domains, including sociology, mathematics, computer science, and physics \cite{de2018physical,newman2018,gleich2014,langville2006}. 

Researchers have developed increasingly comprehensive network representations and analyses \cite{nonlinear2020,renaud2019} to help with data integration and  a variety of applications. 
A prominent example is the generalization of graphs to \emph{multilayer networks} \cite{kivela2014,bianconi2018,yamir2018,whatis2018}, and there have been many efforts to extend centrality measures to multiplex and temporal networks \cite{Halu_Mondragon_Panzarasa_Bianconi_2013,sola2013eigenvector,sole2014centrality,tang2010,alsayed2015,pan2011,kim2012b,williams2015,lerman2010centrality,motegi2012,grindrod2011communicability,estrada2013,Grindrod_Higham_2014,taro2015,rocha2014,rossi2012,you2015distributed,taylor2017eigenvector}. Multilayer network centralities have been used in the study of diverse applications, including social networks \cite{magnani2011ml,magnani2013combinatorial,Halu_Mondragon_Panzarasa_Bianconi_2013,coscia2013you,chakraborty2016cross}, transportation systems \cite{de2015ranking,tavassoli2016most,iacovacci2016functional,tudisco2018node}, economic systems \cite{deford2017multiplex,deford2017new,bardoscia2018multiplex}, neural systems \cite{bassett2013task,de2016mapping,iacovacci2016functional,yu2017selective}, and signal processing of geological time series \cite{lotfi2018centrality}. Moreover, many techniques in centrality analysis are closely connected to the study of various dynamical processes (including in multilayer networks),
such as random walks \cite{gleich2014,masuda2017,ng2011multirank,gomez2013diffusion,Halu_Mondragon_Panzarasa_Bianconi_2013,ding2018centrality}, information spreading \cite{coscia2013you,reiffers2017opinion}, and congestion \cite{de2015ranking}.

\begin{figure}[t]
\centering
\includegraphics[width=.97\linewidth]{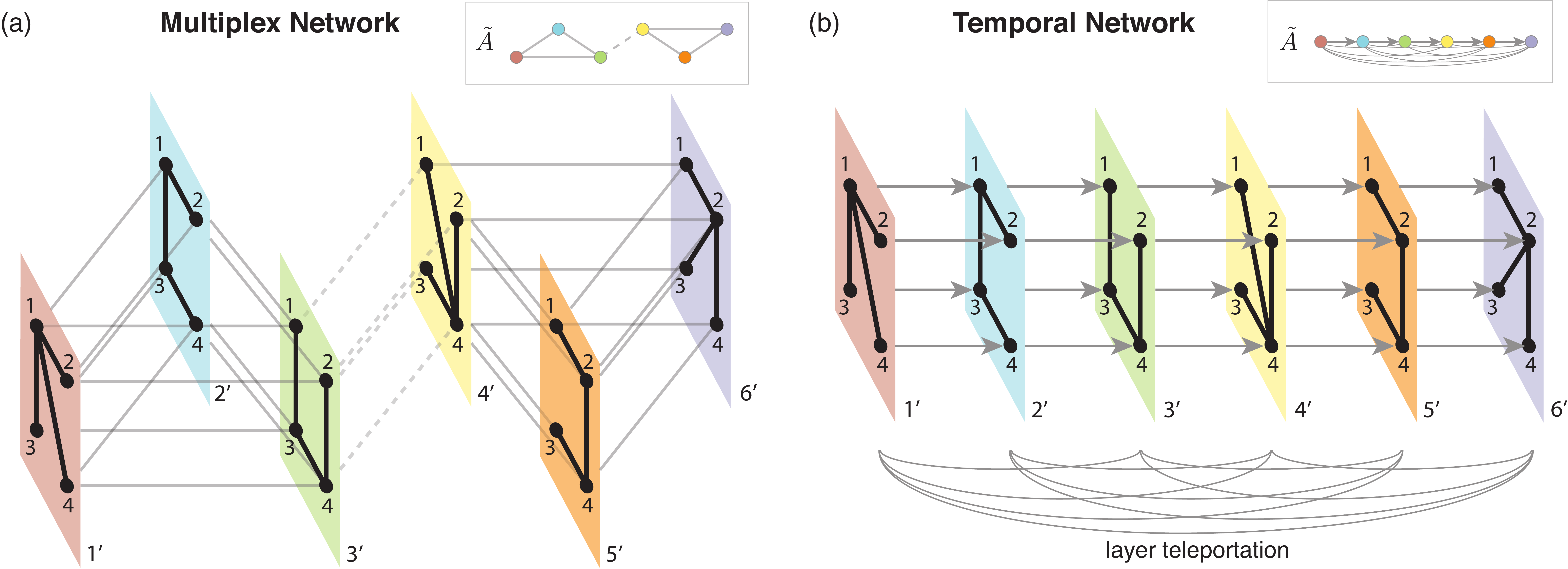}
\vspace{-.3cm}
\caption{ 
{\bf Schematics of two types of multilayer networks.} 
(a)~A multiplex network, in which layers are coupled categorically. 
(b)~A multiplex representation of a discrete-time temporal network, where we couple the sequence of layers through a directed (time-respecting) chain with ``layer teleportation". (See Sec.~\ref{sec:temp} for a definition.) Each inset depicts the interlayer-coupling topology, which we encode (along with interlayer edge weights) in an interlayer-adjacency matrix $\tilde{\bm{A}}$. We assume that the interlayer couplings are ``diagonal'' and ``uniform'' (see Sec.~\ref{sec:back_Multi}), and we take their weights to be $\omega\ge0$. As we illustrate in panels (a) and (b), interlayer coupling can be either undirected or directed. The dashed gray lines between layers 3' and 4' in panel (a) highlight the fact that those edge weights may differ from those of the solid gray lines.
}
\label{fig:toy1}
\end{figure}

We consider two types of multilayer networks (see Fig.~\ref{fig:toy1}): (1) \emph{multiplex networks}, in which layers represent different types of relationships; and (2) \emph{temporal networks}, in which layers represent different time instances or time periods.
We extend the mathematical framework of \emph{supracentrality} matrices, which we developed recently \cite{taylor2017eigenvector} to generalize eigenvector-based centralities (e.g., PageRank \cite{Brin1998conf,pagerank,gleich2014}, eigenvector centrality \cite{bonacich1972}, and hub and authority scores \cite{kleinberg1999}) to multilayer representations of discrete-time temporal networks. Our approach involves coupling centrality matrices that are associated with  individual layers into a larger supracentrality matrix and studying its dominant eigenvector\footnote{{Technically, we study the eigenvector that is associated with the largest positive eigenvalue $\lambda_{\textrm{max}}$ of an irreducible nonnegative matrix. Because the other eigenvalues have magnitudes that are less than or equal to $\lambda_{\textrm{max}}$, we refer to this eigenvalue and its eigenvectors as ``dominant''.}} to obtain \emph{joint}, \emph{marginal}, and \emph{conditional} centralities (see Sec.~\ref{sec:joint}) to quantify the importances of nodes, layers, and node-layer pairs. In this article, we generalize the supracentrality framework of \cite{taylor2017eigenvector} to multiplex networks, which integrate data sets that encode different types of relationships by coupling them as layers of a single multilayer network.

Generalizing centrality measures to multiplex networks and temporal networks are active areas of research \cite{kivela2014,holme2011,holme2015,bianconi2018,yamir2018} (see our discussion in Sec.~\ref{sec:background}), and our supracentrality framework is relevant for such efforts.\footnote{In principle, one can also use supracentrality matrices of higher dimensionality to study networks that are both multiplex and temporal, but we do not study any such examples in this paper.} Our original formulation of supracentrality in \cite{taylor2017eigenvector} focused on temporal networks (see \cite{taylor2019supracentrality} for our more recent work), and it assumed a specific type of multilayer representation with adjacent-in-time coupling. We now extend supracentrality matrices to a broader class of multilayer networks by coupling layers via an interlayer-adjacency matrix $\tilde{\bm{A}}$, where $\tilde{{A}}_{tt'}\ge 0$ encodes the (possibly asymmetric) coupling between layers $t$ and $t'$. We assume ``diagonal'' interlayer coupling (see Sec.~\ref{sec:back_Multi} and \cite{kivela2014}), as we only connect instantiations of the same entity (i.e., node) across different layers. We also assume that the interlayer coupling is ``uniform'', so all edges between layers $t$ and $t'$ have the same weight $\omega \tilde{A}_{tt'}\ge0$. Multilayer networks with both diagonal and uniform interlayer coupling are said to be ``layer-coupled'' \cite{kivela2014}. The value of $\omega$ determines how strongly the layers influence each other. We will show that choices for $\tilde{\bm{A}}$ and $\omega$ significantly affect supracentralities and are useful ``tuning knobs'' to consider when calculating and interpreting supracentralities.

To gain insight into the effects of $\tilde{\bm{A}}$ and $\omega$, we use singular perturbation theory to analyze the dominant eigenspace of supracentrality matrices in the limits of weak ($\omega\to 0^+$) and strong ($\omega\to \infty$) coupling. We show that these limits yield layer decoupling and a type of layer aggregation (which is a form of data fusion), respectively. There are many scenarios in which one couples matrices into a larger ``supramatrix'' --- including the detection of multilayer community structure using a supramodularity matrix \cite{mucha2010,weir2017post} and the study of random walks and diffusion on multilayer networks via supra-Laplacian matrices \cite{gomez2013diffusion,arenas-natphys2013} --- and our perturbative approach reveals insights about the utility of matrix coupling as a general technique for multimodal data integration. Specifically, the singular perturbation theory of Sec.~\ref{sec:limiting} makes no explicit assumption that the block-diagonal matrices are centrality matrices, so our results also characterize the dominant eigenspaces of layer-coupled matrices in other applications, including ones that are unrelated to networks.
 
Our results in Sec.~\ref{sec:limiting} characterize the decoupling and aggregation limits of supracentrality matrices. We illustrate that the limiting dominant eigenspace of a supracentrality matrix depends on a complicated interplay between many factors, including (i) the dominant eigenvectors of the centrality matrix of each layer; (ii) the dominant eigenvectors of the interlayer-adjacency matrix; and (iii) the spectral radii of the layers' centrality matrices. For the $\omega\to\infty$ limit, the dominant eigenspace of a supracentrality matrix depends on a weighted average of the layers' centrality matrices, with weights that are related to the dominant eigenspace of $\tilde{\bm{A}}$. A key factor for the $\omega\to0^+$ limit is whether the layers' individual centrality matrices have identical or different spectral radii. In the latter scenario, we identify and characterize an eigenvector-localization phenomenon in which one or more layers dominate the decoupling limit. Our layer-aggregation and decoupling limits are reminiscent of prior research on supra-Laplacian matrices \cite{sole2013spectral,gomez2013diffusion}, but our coupling matrices and qualitative results both differ from such prior work.

We illustrate our framework with applications to two empirical, multimodal network data sets. First, we study the importances of European airports in a multiplex network in which layers represent different airlines \cite{cardillo2013emergence}. We find, for example, that supracentralities in the weak-coupling limit are dominated by the Ryanair layer; among all layers, this one has the most edges, and its associated adjacency matrix has the largest spectral radius. For intermediate coupling strengths, we observe a  centrality ``boost'' (i.e., an increase relative to the centralities of other nodes) for airports that are central {both} with respect to the Ryanair layer and with respect to a network that is associated with an aggregation of the network layers (specifically, the one from summing the layers' adjacency matrices). We study these phenomena by comparing marginal node centralities to the nodes' intralayer degrees and total degrees (which quantify, respectively, the number of edges of a node in each layer and its total number of edges across all layers). Our second focal example, which we construct using data from the Mathematics Genealogy Project \cite{mgp,MGP_data}, is a temporal network that encodes the graduation and hiring of mathematicians at mathematical-science Ph.D. programs in the United States \cite{myers2011mgp,taylor2017eigenvector}.  Extending \cite{taylor2017eigenvector}, we explore the effects of causality by implementing time-directed interlayer coupling along with \emph{layer teleportation} (see Fig.~\ref{fig:toy1}(b) and our discussion in Sec.~\ref{sec:temp}), which we define analogously to \emph{node teleportation} in PageRank \cite{gleich2014}. (See our recent book chapter \cite{taylor2019supracentrality} for further exploration of layer teleportation.) 
Similar to previous findings for causality-respecting centralities \cite{fenu2015,Grindrod_Higham_2013}, our approach boosts the centralities of node-layer pairs whose edges occur earlier in time (allowing them to causally influence more nodes). In the present paper, this phenomenon manifests as a boost in marginal layer centrality for older time layers. Our numerical experiments highlight the importance of exploring a diverse set of interlayer-coupling architectures $\tilde{\bm{A}}$ and strengths $\omega$ to identify application-appropriate parameter choices.

Our paper proceeds as follows. In Sec.~\ref{sec:back}, we present background information on eigenvector-based centralities, multiplex and temporal networks, and generalizing centralities for such networks. In Sec.~\ref{sec:method}, we present our supracentrality framework. In Sec.~\ref{sec:limiting}, we analyze the weak-coupling and strong-coupling limits. In Sec.~\ref{sec:data}, we study the two empirical data sets. We conclude in Sec.~\ref{sec:discuss} and give the proofs for our main mathematical results in appendices. We present further numerical investigations in Supplementary Materials.

\section{Background Information}\label{sec:back}
%

We now discuss eigenvector-based centralities in Sec.~\ref{sec:back_eig}, multiplex and temporal networks in Sec.~\ref{sec:back_Multi}, and extensions of eigenvector-based centralities for multiplex and temporal networks in Sec.~\ref{sec:background}.

\subsection{Eigenvector-Based Centrality Measures}\label{sec:back_eig}
%
We start with a definition.

\begin{definition}[Monolayer Network] 
Let $\mathcal{G}(\mathcal{V},\mathcal{E})$ be a \emph{monolayer (i.e., single-layer) network} with nodes $\mathcal{V}=\{1,2,\dots,N\}$ and a set $\mathcal{E}\subset \mathcal{V}\times\mathcal{V}\times \mathbb{R}_+$ of positively-weighted edges, where $(i,j,w_{ij})\in\mathcal{E}$ if and only if there exists an edge from $i$ to $j$ with weight $w_{ij}$.  We also encode this network (which is a weighted graph) by an $N\times N$ adjacency matrix $\mathbf{A}$ with entries ${A}_{ij} =w_{ij}$ if $(i,j,w_{ij})\in\mathcal{E} $ and ${A}_{ij}=0$ otherwise. 
\end{definition}


One of the most popular approaches for quantifying the importances of nodes $\mathcal{V}$ in a network is to calculate the dominant eigenvector of a network-related matrix and interpret the eigenvector's entries as a proxy for importance \cite{newman2018,gleich2014}.

\begin{definition}[Eigenvector-Based Centrality] \label{def:EigenBased}
Let $\mathbf{C}=C(\mathbf{A})$ be a ``centrality matrix'' that we obtain via some function $C:\mathbb{R}^{N\times N}\mapsto \mathbb{R}^{N\times N}$ of the adjacency matrix $\mathbf{A}$ for a network $\mathcal{G}(\mathcal{V},\mathcal{E})$. Consider the right 
eigenvector $\mathbf{u}$, which is the solution to
\begin{equation}\label{eq:eigen_1}
	\mathbf{C} \mathbf{u} = \lambda_{\rm{max}} \mathbf{u} \,,
\end{equation}
where $\lambda_{\rm{max}} \in\mathbb{R}$ is the largest {positive} eigenvalue of $\mathbf{C}$. 
Each entry $u_i$ specifies the \emph{eigenvector-based centrality} that is associated with the function $C$ for node $i\in\mathcal{V}$.
{We refer to this eigenvalue and its associated eigenvectors as ``dominant''.}

\end{definition}

Different choices for $C(\mathbf{A})$ yield different notions of centrality, and some are more useful than others. The following are among the most popular eigenvector-based centralities.

\begin{definition}[Eigenvector Centrality \cite{bonacich1972}]\label{def:Eigen}
With the choice $\mathbf{C}^{(EC)} = \mathbf{A}$, Eq.~\eqref{eq:eigen_1} yields \emph{eigenvector centralities} $\{u_i^{(EC)}\}$.
\end{definition}

\begin{definition}[Hub and Authority Scores \cite{kleinberg1999}]\label{def:Hits}
With the choices $\mathbf{C}^{(HS)} = \mathbf{A}\mathbf{A}^\new{*}$ and $\mathbf{C}^{(AS)} = \mathbf{A}^\new{*}\mathbf{A}$, Eq.~\eqref{eq:eigen_1} yields \emph{hub scores} $\{u_i^{(HS)}\}$ and \emph{authority scores} $\{u_i^{(AS)}\}$, respectively.  \new{The symbol $*$ denotes the transpose operator\footnote{\label{foot:transpose}\new{In this paper, we  define $T$ to be the number of layers, so we use the notation $*$ for the transpose operator to avoid confusion. The symbol $*$ is often used to indicate the conjugate transpose operator. This is also a correct interpretation in the present paper, because the matrices that we study have real-valued entries.}}.}
\end{definition}

\begin{remark}
Hub scores and authority scores are, respectively, the  left and right dominant singular vectors of $\mathbf{A}$.
\end{remark}

\begin{definition}[PageRank \cite{pagerank,gleich2014}]\label{def:PageRank}
Consider the choice $\mathbf{C}^{(PR)} = {\sigma}  {(\mathbf{D}^{-1} \mathbf{A})^\new{*}}+ (1-{\sigma})N^{-1}\mathbf{1}\mathbf{1}^\new{*} $, where $\mathbf{D}$ is the diagonal matrix with entries $D_{ii} =  \sum_j A_{ij}$.
%
The quantity {${\sigma}\in[0,1)$} is a ``node'' teleportation parameter (we will assume that ${\sigma}=0.85$ in this paper), and $\mathbf{1}$ is a length-$N$ vector of ones (such that $\mathbf{1}\mathbf{1}^\new{*} $ is an $N\times N$ matrix of ones). Using this choice in Eq.~\eqref{eq:eigen_1}, we obtain \emph{PageRank} centralities $\{u_i^{(PR)}\}$.  
\end{definition}

\new{\begin{remark}
Our definition of PageRank assumes that each edge has at least one out-edge. If there exist ``dangling nodes'' (which lack out-edges), one can 
add self-edges (i.e., either just to such nodes or to all the nodes).
\end{remark}}

\begin{remark}
It is also common to compute PageRank centralities from a left eigenvector \cite{gleich2014}. In the present paper, we use a right-eigenvector formulation to be consistent with the other eigenvector-based centralities. One can recover the other formulation by taking the transpose of  {the matrix $\mathbf{C}$ in} Eq.~\eqref{eq:eigen_1}.
\end{remark}

\begin{remark}
There are other possible teleportation strategies for PageRank, such as ones with a local bias or emphasis on other features \cite{gleich2014}. In such cases, one  {replaces the matrix $\mathbf{1}\mathbf{1}^\new{*} $ by $\mathbf{u}\mathbf{1}^\new{*}$, where the vector $\mathbf{u}$ encodes the biases.}
\end{remark}

In most applications, it is important to choose the function $C$ to ensure that centralities are unique and strictly positive. It is common to use the following two theorems to guarantee these important features.

\begin{theorem}[Perron--Frobenius Theorem for Nonnegative Matrices \cite{bapat1997}] \label{thm:PF}
Let $\mathbf{C}\in\mathbb{R}^{N\times N}$ 
be an irreducible square matrix with nonnegative entries.
It follows that $\mathbf{C}$ has a simple largest positive eigenvalue $\lambda_{\rm{max}}$ and that its right and left eigenvectors are positive and unique. Moreover, if $\mathbf{C}$ is aperiodic,
then 
{$\lambda_{\rm{max}}>|\lambda_i|$ for any $\lambda_i \not=\lambda_{\rm{max}}$.}
\end{theorem}

\begin{theorem}[{Equivalence of} Strong Connectivity {and} Irreducibility \cite{bapat1997}] \label{thm:SC_implies}
Consider the (possibly weighted and directed) network that is associated with a nonnegative square matrix $\mathbf{C}$. {(That is, $\mathbf{C}$ is the adjacency matrix of the associated network.) The matrix $\bf{C}$ is irreducible if and only if} the associated network is strongly connected (i.e., if and only if there exists a path from any origin node to any destination node).
\end{theorem}

One typically seeks a centrality matrix that is irreducible (or, equivalently, a matrix for which the associated network that defined by the weighted edges $\{(i,j,C_{ij})\}$ is strongly connected). Ensuring irreducibility is an important consideration when introducing new types of centrality (including ones with both positive and negative edges \cite{everett2014}). For example, the term $(1-\alpha)N^{-1}\mathbf{1}\mathbf{1}^\new{*} $ in Definition~\ref{def:PageRank} implies that ${ \bf C}^{(PR)}$ is positive (i.e., ${C}^{(PR)}_{ij}>0$ for every $i,j\in\mathcal{V}$), which ensures that ${\bf C}^{(PR)}$ is irreducible, regardless of whether the network with adjacency matrix $\mathbf{A}$ is strongly connected \cite{gleich2014}.

Before continuing, we highlight an eigenvector-based centrality measure that uses both right and left eigenvectors and therefore does not exactly fit Definition~\ref{def:EigenBased}. 
One defines the so-called \emph{dynamical importance} of a node in terms of the change of the leading eigenvalue of ${\bf A}$ under removal of that node from the associated network \cite{restrepo2006} (see also \cite{tong2012gelling}). In practice, as shown in \cite{restrepo2006}, one can approximate dynamical importance to first order (provided one does not lose strong connectivity when removing the node) with an expression that depends on the leading right and left eigenvectors of ${\bf A}$. Other eigenvector-based centralities that involve two or more eigenvectors that one obtains through matrix perturbations have been developed to cater to particular applications, including the spread of infectious diseases \cite{taylor2011,taylor2012}, percolation \cite{restrepo2006,taylor2012network}, and synchronization 
\cite{taylor2016synchronization}. Such analysis of perturbations of dynamical systems on networks is also related to notions of eigenvalue and eigenvector elasticities \cite{kampmann1996feedback,kampmann2006loop,gonccalves2009behavior}.

\subsection{Multiplex and Temporal Networks}\label{sec:back_Multi}
%
The different layers of a multilayer network can encode different types of connections and/or interacting systems \cite{kivela2014}, including interconnected infrastructures \cite{haimes2001leontief}, categories of social ties \cite{krackhardt1987cognitive}, networks at different instances of time \cite{valdano2015analytical}, and many others. By considering the various possibilities for interactions between nodes within and across layers, one can obtain a taxonomy for different types of multilayer networks \cite{kivela2014}. We focus on two popular situations: \emph{multiplex networks}, in which different layers represent different types of interactions; and \emph{temporal networks}, in which layers represent different time instances or time periods. 

We give formal definitions that are salient for multiplex and temporal networks. For both types of multilayer networks, it is convenient to refer to a node $i$ in a layer $t$ as a \emph{node-layer pair} $(i,t)$.

\begin{definition}[Uniformly and Diagonally Coupled Multiplex Network]\label{def:mux}
Let $\mathcal{G}(\mathcal{V},\{\mathcal{E}^{(t)}\},\tilde{\mathcal{E}})$ be a multilayer network with nodes $\mathcal{V}=\{1,\dots,N\}$ and $T$ layers, with interactions between node-layer pairs encoded by the sets $\{\mathcal{E}^{(t)}\}$ of weighted edges, where $(i,j,w^t_{ij})\in\mathcal{E}^{(t)}$ if and only if there is an edge $(i,j)$ with weight $w^t_{ij}$ in layer $t$. The set $\tilde{\mathcal{E}}=\{(s,t,\tilde{w}_{st})\}$ encodes the topology and weights for coupling separate instances of the same node between a layer pair $(s,t)\in\{1,\dots, T\}\times \{1,\dots, T\}$. Equivalently, one can encode a multiplex network as a set $\{\mathbf{A}^{(t)}\}$ of adjacency matrices, such that $ {A}_{ij}^{(t)}=w^t_{ij}$ if $(i,j,w^t_{ij})\in\mathcal{E}^{(t)}$ and $ {A}_{ij}^{(t)}=0$ otherwise, along with an interlayer-adjacency matrix $\tilde{\bf{A}}$ with components $\tilde{A}_{st} = \tilde{w}_{st}$ if $(s,t,\tilde{w}_{st}) \in \tilde{\mathcal{E}}$ and $\tilde{A}_{ij}^{(t)}=0$ otherwise.
\end{definition}

See Fig.~\ref{fig:toy1}(a) for a pedagogical example of a small multiplex network. The multiplex coupling in Definition \ref{def:mux} is ``diagonal'' because we only allow coupling between a node in one layer and that same node in other layers, and it is ``uniform'' because the coupling between two layers is identical for all nodes in those two layers. A multilayer network with both of these conditions is called ``layer-coupled'' \cite{kivela2014}. Our choice to represent interlayer couplings via $\tilde{\bf{A}}$ is a generalization of the special, but common, case in which the interlayer edge weights are identical for all layer pairs (i.e., $\tilde\omega_{st}=\omega$ for all $s$ and $t$).
Although there are many other coupling strategies \cite{boccaletti2014,kivela2014}, we focus on uniform and diagonal coupling because it is one of the simplest and most popular coupling schemes. The interlayer-adjacency matrix $\tilde{\bf{A}}$ already allows a great deal of flexibility, and (as we will describe in Sec.~\ref{sec:method}) these restrictions impose matrix symmetries that we can exploit to derive results when the layers are coupled either very weakly or very strongly. 

We use a similar multilayer network representation to study temporal networks.

\begin{definition}[Discrete-Time Temporal Network]\label{def:time}
A \emph{discrete-time temporal network} consists of nodes $\mathcal{V}=\{1,\dots,N\}$ and a sequence of network layers. We denote the network either by $\mathcal{G}(\mathcal{V},\{\mathcal{E}^{(t)}\})$ or by a sequence $\{\mathbf{A}^{(t)}\}$ of adjacency matrices such that ${A}_{ij}^{(t)}=w^t_{ij}$ if $(i,j,w^t_{ij})\in\mathcal{E}^{(t)}$ and $ {A}_{ij}^{(t)}=0$ otherwise. 
\end{definition}

Definition \ref{def:time} makes no explicit assumptions about how the layers are coupled. That is, a discrete-time temporal network consists of a set of nodes and an ordered set of layers. It is common, however, for temporal networks to also include coupling between layers, such as by representing them (as we do) as a multiplex network with a diagonal interlayer coupling that respects the arrow of time.

\subsection{Extensions of Centrality for Multiplex and Temporal Networks}\label{sec:background}
%
There has been a recent explosion of research on centrality measures for multilayer networks \cite{kivela2014,yamir2018}. Much of this work is related to work on generalizing network properties such as node degree \cite{magnani2011ml,de2013b,battiston2014structural,tavassoli2016most} and shortest paths, with the latter leading to generalizations of notions like betweenness centrality \cite{magnani2013combinatorial,sole2014centrality,sole2016random,chakraborty2016cross} and closeness centrality \cite{magnani2011ml,sole2016random}. Of particular relevance to the present paper are generalizations of eigenvector-based centralities to multiplex networks. Salient notions that have been generalized include eigenvector centrality \cite{de2013b,sola2013eigenvector,battiston2014structural,deford2017new,deford2017multiplex}, hub and authority scores \cite{kolda2006tophits,spatocco2018new,rahmede2017centralities,tudisco2018node}, and PageRank \cite{ng2011multirank,Halu_Mondragon_Panzarasa_Bianconi_2013,ding2018centrality}.
These extensions have employed various strategies; we briefly discuss several of them. 

One strategy is to represent a multiplex network as a tensor and use tensor decompositions \cite{kolda2006tophits,wu2019tensor}. Another strategy is to define a system of centrality dependencies in which large-centrality elements (nodes, layers, and so on) connect to other large-centrality elements, and one simultaneously solves for multiple types of centrality \cite{rahmede2017centralities,spatocco2018new,tudisco2018node} as a fixed-point solution of the system of (possibly nonlinear) dependencies. For example, \cite{rahmede2017centralities} and \cite{tudisco2018node} defined centralities for both nodes and layers such that highly-ranked layers have highly-ranked nodes and highly-ranked nodes are in highly-ranked layers. 
In a third strategy, which is the one that most closely resembles our present approach, one constructs a \emph{supramatrix} of size $NT \times NT$ for $N$ nodes and $T$ layers, such that the dominant eigenvector of the supramatrix gives the centralities of the node-layer pairs $\{(i,t)\}$. For example, \cite{sola2013eigenvector,romance2015perron,deford2017multiplex,deford2017new} constructed a supramatrix by using the Khatri--Rao matrix product between a matrix that encodes interlayer connections and a block matrix that has the layers' adjacency matrices as its \new{block} columns. 
Another approach involves computing one or more centralities independently for each layer and then using a consensus ranking \cite{posfai2019consensus}. One can also generalize monolayer centrality measures to construct multilayer ``versatility'' measures \cite{de2015ranking}.

There have also been many efforts to generalize centralities to temporal networks. There are extensive discussions of such efforts in \cite{taylor2017eigenvector} and \cite{liao2017ranking}. To add to these lists, we briefly highlight several contributions that appeared recently or were not mentioned in \cite{taylor2017eigenvector}. Arrigo and Higham \cite{arrigo2017sparse} introduced a method to efficiently estimate temporal communicability (a generalization of Katz centrality), which has been applied to a variety of applications, including in neuroscience \cite{mantzaris2013dynamic} and the spread of infectious diseases \cite{mantzaris2013dynamic2,chen2016dynamic}. Huang and Yu~\cite{huang2017dynamic} extended a measure called dynamic-sensitive centrality to temporal networks. References \cite{praprotnik2015spectral,huang2017centrality,flores2018eigenvector} introduced variants of eigenvector centrality for temporal networks. We highlight \cite{flores2018eigenvector} in particular, because it explored connections between continuous and discrete-time calculations of temporal centralities. Nathan et al.~\cite{nathan2017ranking} introduced an efficient algorithm for computing a centrality for streaming graphs. Although methods for streaming and continuous-time networks are important (see, e.g., \cite{walid2018} for a generalization of PageRank to such situations), we restrict our attention to discrete-time temporal networks, especially because we seek to further bridge the literatures on temporal and multiplex networks.


In Sec.~\ref{sec:method}, we introduce a new construction, which is based on a Kronecker product, for a supracentrality matrix. This construction generalizes our previous work \cite{taylor2017eigenvector}, where we introduced a supracentrality framework for temporal networks and assumed adjacent-in-time coupling (i.e., that $\tilde{A}_{tt'}=1$ for $|t-t'|=1$ and $\tilde{A}_{tt'}=0$ otherwise). Our new formulation introduces an interlayer-adjacency matrix $\tilde{\bf{A}}$, allowing our framework to flexibly cater to either multiplex or temporal networks. 
Note that there are also multilayer representations of temporal networks that are not multiplex. One such approach is to connect  node-layer pair $(i,t)$ to $\{(j,t+1)\}$ for $j\in\{j:A^{(t)}_{ij}\not=0\}$, yielding a supra-adjacency matrix with identity matrices on the block diagonal and the layers' adjacency matrices on the off-diagonal blocks that lie directly above the diagonal blocks. 
These interlayer edges are nondiagonal, because they connect nodes in one layer to different nodes in a neighboring layer. This formulation, which is connected mathematically \cite{fenu2015} to matrix-iteration-based centrality measures for temporal networks \cite{Grindrod_Higham_2013,Grindrod_Higham_2014,grindrod2011communicability}, has been used to study time-dependent functional brain networks \cite{muldoon2018}.
Additionally, multilayer networks with non-diagonal edges were used in \cite{valdano2015analytical} to study disease spreading on temporal networks. One can also to choose to study multiplex and temporal networks without interlayer coupling by independently considering each layer in isolation.

\begin{table}[h]
\caption{Summary of our mathematical notation for objects of different dimensions.}
\small{\centering
\begin{tabular}{c c c} 
\hline\hline 
Typeface & Class & Dimension  \\ [0.5ex]
\hline 
$\mathbb{M}$ & matrix & $NT\times NT$   \\ 
$\mathbf{M}$ & matrix & $N\times N$  \\
$\bm{M}$ & matrix & $T\times T$   \\
$\mathbbm{v}$ & vector & $NT\times 1$  \\
$\mathbf{v}$ & vector & $N\times 1$   \\ 
$\bm{v}$ & vector & $T\times 1$   \\ 
$M_{ij}$ & scalar & 1   \\ 
$v_i$ & scalar & 1   \\
[1ex] 
\hline 
\end{tabular}\\
\label{table:notation}} 
\end{table}

\section{Supracentrality Framework for Multiplex and Temporal Networks}\label{sec:method}
%
We now present a supracentrality framework that provides a common mathematical foundation for eigenvector-based centralities for layer-coupled multiplex and temporal networks. In Sec.~\ref{sec:supra}, we define supracentrality matrices. In Sec.~\ref{sec:joint}, we define joint, marginal, and conditional centralities; we prove their uniqueness and positivity under certain conditions. In Sec.~\ref{sec:peda}, we give a pedagogical example to illustrate these types of centralities. 

We summarize our key mathematical notation in Table~\ref{table:notation}. We use the subscripts $\new{i,j}\in \mathcal{V}$ to enumerate nodes, the subscripts $\new{s,t}\in\mathcal\{1,\dots,T\}$ to enumerate layers, and the subscripts $\new{p,q}\in\mathcal\{1,\dots,NT\}$ to enumerate node-layer pairs. 

\subsection{Supracentrality Matrices}\label{sec:supra}
%
We first define a supracentrality matrix, in a way that generalizes the definition in \cite{taylor2017eigenvector}, for networks that are either multiplex or temporal.

\begin{definition}[Supracentrality Matrix]\label{def:Supracentrality}
Let $\{{\bf C}^{(t)}\}$ be a set of $T$ centrality matrices for a multilayer network whose layers have a common set $\mathcal{V}=\{1,\dots,N\}$ of nodes, and suppose that ${C}^{(t)}_{ij} \ge 0 $. 
Let $\tilde{\bm{A}}$, with $ \tilde{A}_{ij}\ge 0$, be a $T\times T$ interlayer-adjacency matrix that encodes the interlayer couplings.
We define a family of \emph{supracentrality matrices} $\mathbb{C}(\omega)$, which are parameterized by the interlayer-coupling strength $\omega \ge 0$, of the form 
\begin{align}
	{\mathbb{C}} (\omega) &=  \hat{\mathbb{C}} +  \omega \hat{\mathbb{A}}\,
	= \left[ \begin{array}{cccc} 
 \mathbf{C}^{(1)} & {\bf 0} &  {\bf 0}&\dots\\ 
{\bf 0} & \mathbf{C}^{(2)} & {\bf 0}& \dots\\ 
{\bf 0} &  {\bf 0} &  \mathbf{C}^{(3)} &\ddots\\
 \vdots&   \vdots & \ddots&\ddots\\
 \end{array}
 \right] + \omega
 \left[ \begin{array}{cccc} 
  \tilde{A}_{11} \mathbf{I}&  \tilde{A}_{12} \mathbf{I} & \tilde{A}_{13} \mathbf{I}  &\dots \\ 
\tilde{A}_{21} \mathbf{I} & \tilde{A}_{22} \mathbf{I} & \tilde{A}_{23} \mathbf{I}& \dots \\ 
 \tilde{A}_{31} \mathbf{I} & \tilde{A}_{32} \mathbf{I}& \tilde{A}_{33} \mathbf{I} &\dots\\
 \vdots  &   \vdots&\vdots&\ddots\\\\
 \end{array}
 \right] \, , \label{eq:supracentrality}
\end{align}
where $\hat{\mathbb{C}} = \text{\rm diag}[ \mathbf{C}^{(1)},\dots, \mathbf{C}^{(T)}]$ and $\hat{\mathbb{A}}=\tilde{\bm{A}}\otimes \bf{I}$ denotes the Kronecker product of $\tilde{\bm{A}}$ and $\bf{I}$.
\end{definition}

\begin{remark}
For layer $t$, the matrix ${\bf C}^{(t)}$ can be any matrix whose dominant eigenvector is of interest. We focus on centrality matrices, such as those that are associated with eigenvector centrality (see Definition \ref{def:Eigen}), hub and authority scores (see Definition \ref{def:Hits}), and PageRank (see Definition \ref{def:PageRank}). Additionally, one can scale each ${\bf C}^{(t)}$ by a layer-specific weight. (Such a scaling has the potential to benefit both multilayer community detection \cite{pamfil2018} and layer-averaged clique detection \cite{nayar2015improved}.) One can easily incorporate such weighting into Eq.~\eqref{eq:supracentrality} by redefining the centrality matrices $\{{\bf C}^{(t)}\}$ to include the weights.
\end{remark}

The supracentrality matrix $\mathbb{C}(\omega) $ of size $NT\times NT$ encodes the effects of two distinct types of connections: the layer-specific centrality entries $\{ {C}^{(t)}_{ij}\}$ in the diagonal blocks relate centralities between nodes within layer $t$, whereas entries in the off-diagonal blocks encode coupling between layers.  The supramatrix
$\hat{\mathbb{A}}=\tilde{\bm{A}}\otimes \bf{I}$ implements uniform and diagonal coupling: the matrix $\bf{I}$ encodes diagonal coupling, and any two layers $t$ and $t'$ are uniformly coupled, because all interlayer edges between them have the identical weight $w\tilde{A}_{tt'}$. 
The choice of undirected, adjacent-in-time interlayer coupling (i.e., $\tilde{A}_{tt'} = 1$ if $|t-t'|=1$ and $\tilde{A}_{tt'}=0$ otherwise) recovers the supracentrality matrix that we studied in \cite{taylor2017eigenvector}. In the present paper, we generalize that notion of a supracentrality matrix by using an interlayer-adjacency matrix $\tilde{\bm{A}}$, which allows us to implement a wide variety of interlayer coupling topologies. In the context of multiplex networks, we hypothesize that different choices for $\tilde{\bm{A}}$ will have different benefits. In the context of temporal networks, we will study (see Sec.~\ref{sec:temp}) the effects of letting $\tilde{\bm{A}}$ encode a directed, time-respecting chain with ``layer teleportation" (see Eq.~\eqref{eq:tele}). This yields supracentrality results that we will contrast with those in \cite{taylor2017eigenvector}.

\subsection{Joint, Marginal, and Conditional Centralities}\label{sec:joint}
%

~\new{We now study supracentralities in the form of joint, marginal, and conditional centralities.}


The defining feature of eigenvector-based centrality measures is that one computes and studies a dominant eigenvector of a centrality matrix.  
We study the right dominant-eigenvalue equation 
\begin{equation} \label{eq:eig_eq}
	\mathbb{{C}}(\omega)\mathbbm{v}(\omega) = \lambda_{\rm{max}}(\omega)\mathbbm{v}(\omega)\,,
\end{equation}
where we interpret entries in the right dominant eigenvector $\mathbbm{v}(\omega)$ as centrality measures for the  node-layer pairs $\{(i,t)\}$. The vector $\mathbbm{v}(\omega)$ has a block form: its first $N$ entries encode the joint centralities for layer $t=1$, its next $N$ entries encode the joint centralities for layer $t=2$, and so on. Therefore, as we now describe, it can be useful to reshape the block vector $\mathbbm{v}(\omega)$ into a matrix.

Following \cite{taylor2017eigenvector}, we use the concepts of \emph{joint}, \emph{marginal}, and \emph{conditional} centralities to develop our understanding of the importances of nodes and layers from the values of $\mathbbm{v}(\omega)$.

\begin{definition}[Joint Centralities of Node-Layer Pairs \cite{taylor2017eigenvector}]\label{def:joint}
Let $\mathbb{C}(\omega)$ be a supracentrality matrix given by Definition~\ref{def:Supracentrality}, and let $\mathbbm{v}(\omega)$ be its right dominant eigenvector. We encode the \emph{joint centrality} of node $i$ in layer $t$ via the $N\times T$ matrix ${\bf W}(\omega)$ with entries 
\begin{equation}
	W_{it}(\omega)  = \mathbbm{v}_{N(t-1) + i}(\omega)\, . \label{eq:joint2}
\end{equation}
\end{definition}
\begin{remark}
We refer to $W_{it}(\omega)$ as a ``joint centrality'' because it reflects the importance both of node $i$ and of layer $t$.
\end{remark}

\begin{definition}[Marginal Centralities of Nodes and Layers \cite{taylor2017eigenvector}]\label{def:marg}
Let ${\bf W}(\omega) $ encode the joint centralities given by Definition \ref{def:joint}. 
We define the \emph{marginal layer centrality} (MLC) ${x}_{t}(\omega)$ and \emph{marginal node centrality} (MNC) $\hat{x}_{i}(\omega)$
by
\begin{equation}
	 {x}_{t}(\omega) = \sum_{i}W_{it}(\omega) \,, ~~
	 	\hat{x}_{i}(\omega) = \sum_{t} W_{it}(\omega) \, . \label{eq:marg2}  
\end{equation}
\end{definition}

\begin{definition}[Conditional Centralities of Nodes and Layers \cite{taylor2017eigenvector}]\label{def:cond}
Let the set $\{W_{it}(\omega)\}$ be the joint centralities given by Definition \ref{def:joint}, and let $\{{x}_{t}(\omega)  \}$ and $\{\hat{x}_{i}(\omega)  \}$, respectively, be the marginal layer and node centralities from
Definition \ref{def:marg}.  We define the \emph{conditional centralities} of nodes and layers by
\begin{equation}
	Z_{it} (\omega)= W_{it}(\omega)/ {x}_t(\omega)\,, ~~\hat{Z}_{it}(\omega) = W_{it}(\omega)/\hat{x}_i(\omega)\,,
\label{eq:cond}
\end{equation}
where $Z_{it} (\omega)$ gives the centrality of node $i$ conditioned on layer $t$ and $\hat{Z}_{it} (\omega)$ gives the centrality of layer $t$ conditioned on node $i$.
\end{definition}

The quantity $Z_{it}(\omega)$ indicates the importance of node $i$ relative to other nodes in layer $t$. By contrast, the joint node-layer centrality $W_{it}(\omega)$ measures the importance of node-layer pair $(i,t)$ relative to all node-layer pairs. 

We now present a new theorem that ensures the uniqueness and positivity of the above 
supracentralities.

\begin{theorem}[Uniqueness and Positivity of Supracentralities] \label{thm:unique}
Let $\mathbb{C}(\omega)$ be a supracentrality matrix given by Eq.~\eqref{eq:supracentrality}. Additionally, suppose that $\tilde{\bm{A}}$ is an adjacency matrix for a strongly connected graph and that the sum $\sum_t \mathbf{C}^{(t)}$ is an irreducible nonnegative matrix. It then follows that $\mathbb{C}(\omega)$ is irreducible, nonnegative, and has a simple largest positive eigenvalue $\lambda_{\rm{max}}(\omega)$ with corresponding left eigenvector $\mathbbm{u}(\omega)$ and right eigenvector $\mathbbm{v}(\omega)$, which are unique and consist of positive entries. Moreover, the centralities $\{W_{it}(\omega)\}$, $\{x_t(\omega)\}$, $\{\hat{x}_i(\omega)\}$, $ \{Z_{it}(\omega)\}$, and $ \{\hat{Z}_{it}(\omega)\}$ are positive and well-defined. 
\begin{proof}
See Appendix~\ref{app:unique}.
\end{proof}
\end{theorem}
\begin{remark}
If we also assume that $\mathbb{C}(\omega)$ is aperiodic, then it is also true that $\lambda_{\rm{max}}(\omega)$ \new{is larger in magnitude than the other eigenvalues}.
\end{remark}

\begin{figure}[t]
\centering
\includegraphics[width=.45\linewidth]{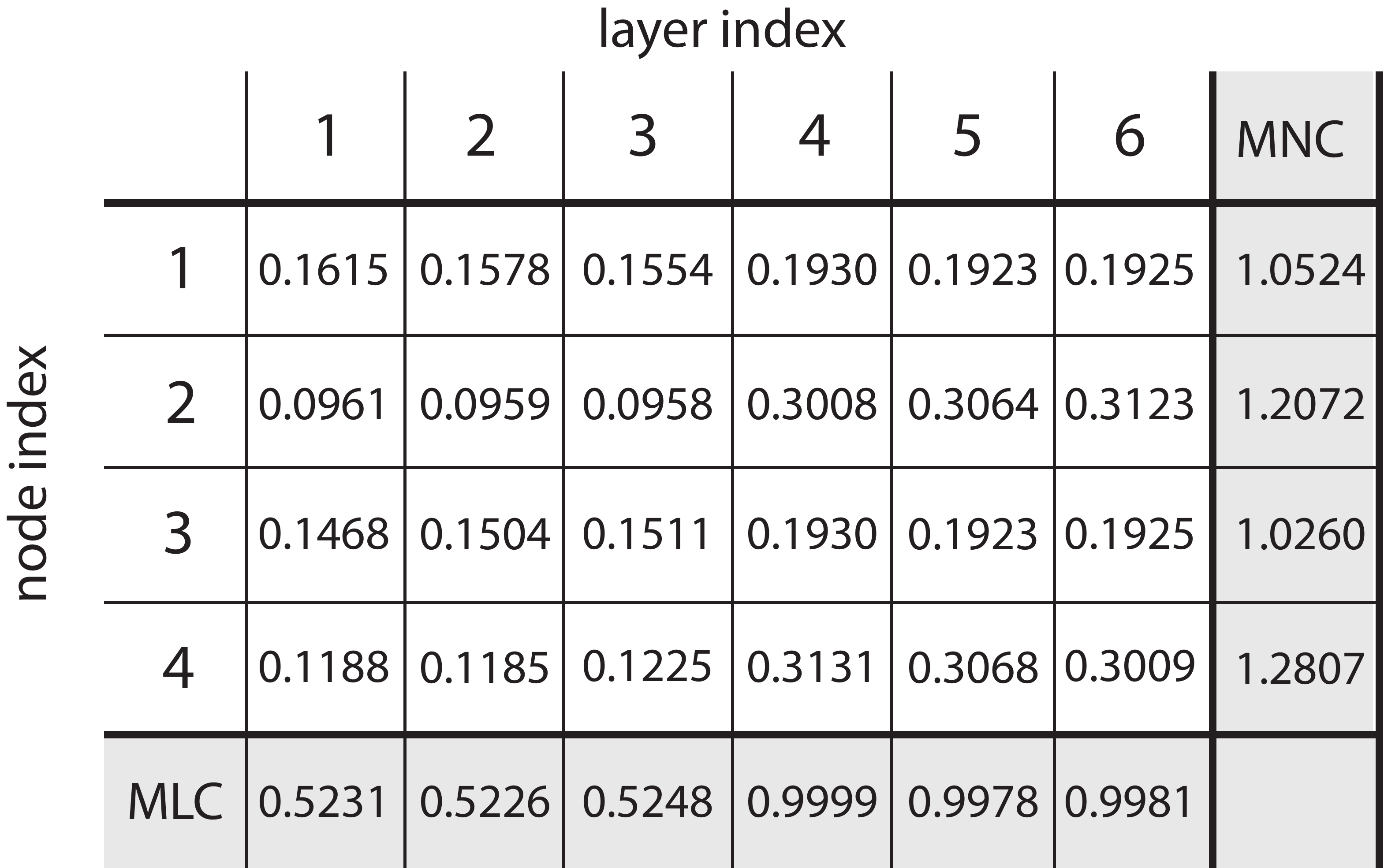}
\caption{Joint node-layer centralities $\{W_{it}(\omega)\}$ given by Definition~\ref{def:joint} (white cells), with corresponding marginal layer centralities (MLCs) $\{{x}_{t}(\omega)\}$ and marginal node centralities (MNCs) $\{\hat{x}_{i}(\omega)\}$ given by Definition~\ref{def:marg} (gray cells), for the multiplex network in Fig.~\ref{fig:toy1}(a). These computations are for diagonally and uniformly coupled (i.e., layer-coupled) eigenvector centralities in which the layers' centrality matrices are given by Definition~\ref{def:Eigen} with $\omega=1$.
}
\label{fig:centralities_toy1}
\end{figure}
 
\subsection{Pedagogical Example that Illustrates Different Coupling Regimes}\label{sec:peda}
%
In Fig.~\ref{fig:centralities_toy1}, we illustrate the concepts of joint and marginal centralities for the multiplex network in Fig.~\ref{fig:toy1}(a). This network has $N=4$ nodes and $T=6$ layers, and we study the interlayer-adjacency matrix $\tilde{\bm{ A}}$ that we showed in the inset of Fig.~\ref{fig:toy1}(a).  We set $\tilde{A}_{tt'}=1$ for all depicted interlayer couplings, except for the coupling of layers $3$ and $4$, for which we set $\tilde{A}_{34}=\tilde{A}_{43}=0.01$. With 
these interlayer edge weights, the interlayer-coupling network that is associated with $\tilde{\bm{ A}}$ has two natural communities of densely-connected nodes. For this experiment (and our other experiments), we typically find that conditional node centralities provide the most useful insights. 

\begin{figure}[t]
\centering
\includegraphics[width=.9\linewidth]{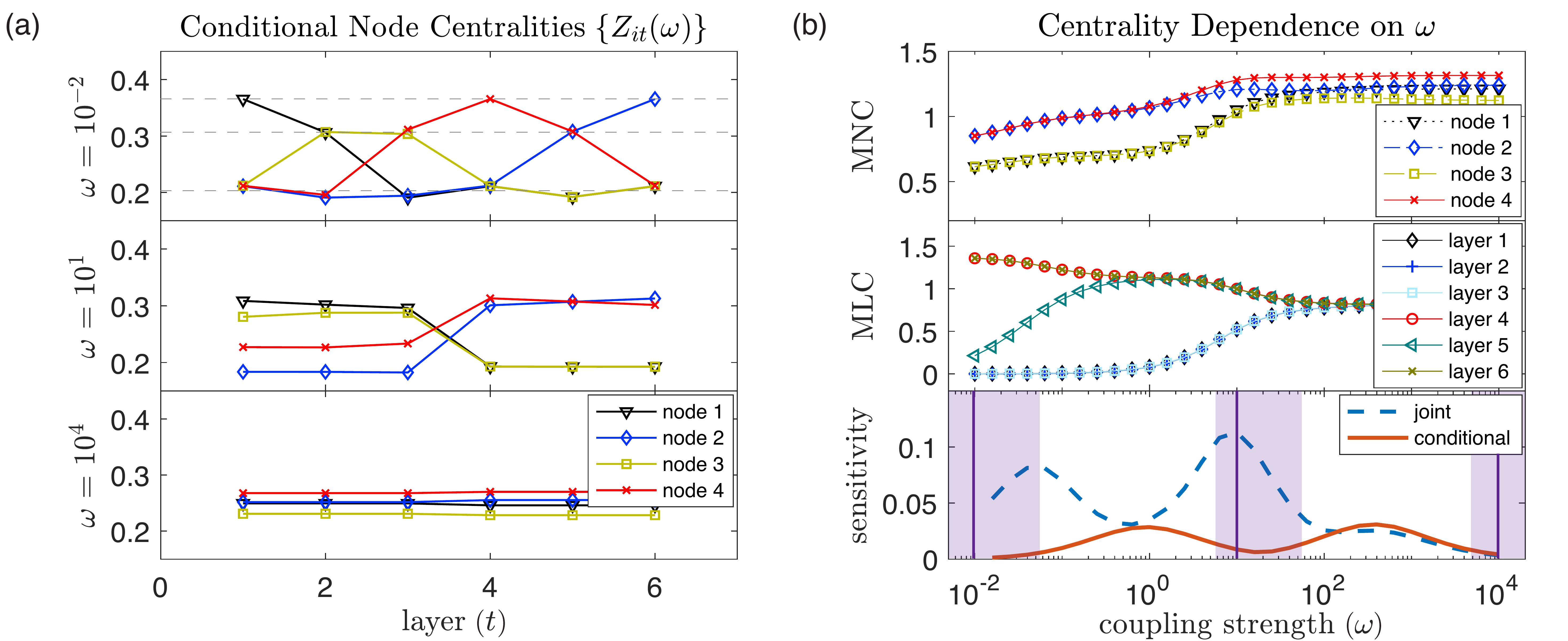}
\vspace{-.3cm}
\caption{
Eigenvector centralities (see Definition~\ref{def:Eigen}) for the uniformly and diagonally coupled multilayer network with interlayer-adjacency matrix $\tilde{\bm{ A}}$ with elements $\tilde{A}_{tt'}=1$, except for the pair $(t,t') = (3,4)$ (for which $\tilde{ A}_{34}=\tilde{A}_{43}=0.01$), and the coupled layers in Fig.~\ref{fig:toy1}(a).
(a) Conditional node centralities $\{Z_{it}(\omega)\}$ versus layer $t$ for interlayer-coupling strengths $\omega \in\{10^{-2},10,10^4\}$. 
{The horizontal dashed lines in the top panel indicate the intralayer node degrees ($d_i^{(t)}=\sum_j A_{ij}^{(t)}$), which  we normalize by multiplying each $d_i^{(t)}$ by   $(\sum_i d_i^{(t)})^{-1/2}$}.  The conditional node centralities are 
$Z_{it}(\omega)\approx 0.3656$ for $d_i^{(t)}=3$,
$Z_{it}(\omega)\approx 0.3096$ for $d_i^{(t)}=2$, and
$Z_{it}(\omega)\approx 0.2021$ for $d_i^{(t)}=1$.
(b) MNC and MLC versus $\omega$ on the interval $[10^{-2},10^{4}]$, which we discretize and use the values $\omega_s=10^{-2+0.2s}$ for $s\in\{0,\dots,30\}$. The bottom panel depicts a measure for the sensitivity of centralities with respect to $\omega$. The dashed blue curve indicates the stepwise magnitude of change, $\|{\bf W}(\omega_s) - {\bf W}(\omega_{s-1}) \|_F$, for the joint centralities; the solid red curve indicates the change $\|{\bf Z}(\omega_s) - {\bf Z}(\omega_{s-1}) \|_F$ for the conditional centralities. 
Because the curve for $\|{\bf Z}(\omega_s) - {\bf Z}(\omega_{s-1}) \|_F$ is bimodal, there are three ranges of $\omega$ that are separated by two peaks; we highlight these ranges with the shaded regions. The three vertical lines in the bottom panel indicate the values of $\omega$ from (a).
}
\label{fig:centralities_toy1b}
\end{figure}

 In Fig.~\ref{fig:centralities_toy1b}(a), we plot the conditional centralities $\{Z_{it}(\omega)\}$ of node-layers for three different choices of the interlayer coupling-strength $\omega$.  These choices represent three centrality regimes (which we illustrate in Fig.~\ref{fig:centralities_toy1b}(b)) that we observe by exploring centralities across a range of $\omega$ values. In the top two panels of Fig.~\ref{fig:centralities_toy1b}(b), we plot the MNC and MLC values versus $\omega$. In the bottom panel, we quantify the ``sensitivity'' of the joint and conditional centralities to perturbations of $\omega$. Specifically, we consider $\omega$ in the interval $[10^{-2},10^{4}]$ discretized by $\omega_s=10^{-2+0.2s}$ for $s\in\{0,\dots,30\}$.  We plot the stepwise magnitudes of the changes $\|{\bf W}(\omega_s) - {\bf W}(\omega_{s-1}) \|_F$ of the joint centralities and the changes $\|{\bf Z}(\omega_s) - {\bf Z}(\omega_{s-1}) \|_F$ of the conditional centralities, where $\|\cdot \|_F$ denotes the Frobenius norm. We identify three regimes for which the conditional centralities are robust. (See the shaded regions in the bottom panel of Fig.~\ref{fig:centralities_toy1b}(b).) 
The peaks in Fig.~\ref{fig:centralities_toy1b}(b) indicate values of $\omega$ where conditional centralities are most sensitive to perturbations of $\omega$; other choices for $\omega$ are more robust to a perturbation of $\omega$. Interestingly, the peaks and troughs for the curves for $\|{\bf W}(\omega_s) - {\bf W}(\omega_{s-1}) \|_F$ and $\|{\bf Z}(\omega_s) - {\bf Z}(\omega_{s-1}) \|_F$ do not coincide. We focus on robust values of ${\bf Z}(\omega)$, because we generally find that conditional centralities provide the most interpretable and insightful results among our supracentralities. See Sec.~\ref{sec:joint} 
and Sec.~5 of \cite{taylor2017eigenvector}.

We summarize these three regimes as follows.
{\begin{itemize}
\item {\bf Weak-Coupling Regime.} 
The top panel of Fig.~\ref{fig:centralities_toy1b}(a), which shows results for interlayer-coupling strength $\omega = 10^{-2}$, depicts a regime in which the conditional node centralities resemble the centralities of uncoupled layers as $\omega \rightarrow 0^+$. In this example, we observe that the conditional centralities correlate strongly with intralayer degrees ($d_i^{(t)} = \sum_j A^{(t)}_{ij}$). Specifically, as we indicate with the horizontal dashed lines, the conditional centralities approximately equal one of three values: $Z_{it}\approx 0.3656$ for $d_i^{(t)}=3$, $Z_{it}\approx 0.3096$ for $d_i^{(t)}=2$, and $Z_{it}\approx 0.2021$ for $d_i^{(t)}=1$. We analyze the $\omega \rightarrow 0^+$ limit in Sec.~\ref{sec:weak}.

\item {\bf Strong-Coupling Regime.} 
The bottom panel of Fig.~\ref{fig:centralities_toy1b}(a), which shows results for $\omega = 10^{4}$, depicts a regime in which the conditional centralities approach the centralities of a layer-aggregated centrality matrix as $\omega \rightarrow \infty$.  In this regime, the conditional centralities $Z_{it}$ of each node $i$ limit to a value  $\alpha_i$ that is constant across the layers; that is, $Z_{it}\to \alpha_i$ for all $t$. We analyze the $\omega \rightarrow \infty$ limit in Sec.~\ref{sec:strong}.

\item {\bf Intermediate-Coupling Regime.}  
The middle panel of Fig.~\ref{fig:centralities_toy1b}(a) depicts an intermediate regime. We show conditional centralities at $\omega = 10$, which is a value of $\omega$ that lies between the locations of the two peaks of the bimodal curve $\|{\bf Z}(\omega_s) - {\bf Z}(\omega_{s-1}) \|_F$ that we show in Fig.~\ref{fig:centralities_toy1b}(b). For $\omega = 10$, we observe two subsets of the $Z_{it}$ conditional centrality values: those for layers $t\in\{1,2,3\}$ are very similar to each other, and those for layers $t\in\{4,5,6\}$ are also similar to each other. This pattern arises directly from the layer-coupling scheme in Fig.~\ref{fig:toy1}(a), where these two sets of layers correspond to two communities in the interlayer-adjacency matrix. (Recall that the coupling between layers $3$ and $4$ is $100$ times weaker than the other couplings.)  In Sec.~SM1 of the Supplementary Materials, we show that the curve $\|{\bf Z}(\omega_s) - {\bf Z}(\omega_{s-1}) \|_F$ becomes unimodal as we increase the coupling   between layers $3$ and $4$.
\end{itemize}}

We classify the strong-coupling and weak-coupling regimes by considering whether or not the observed supracentralities are strongly correlated with those of either asymptotic limit (i.e., either $\omega\to0^+$ or $\omega\to\infty$). The intermediate regime arises from an interplay between the topologies and edge weights of  the layers and the interlayer couplings, so these multilayer centralities provide insights that cannot be observed by studying the network layers in isolation or in aggregate. This example also illustrates that it is important  to explore various coupling strengths $\omega$ and various interlayer-adjacency matrices $\tilde{\bm{A}}$ to identify supracentralities that are appropriate for a given application. See \cite{code} for our {\sc Matlab} code that computes supracentralities and reproduces Fig.~\ref{fig:centralities_toy1b} and our other experiments in this paper.

\section{Limiting Behavior for Weak and Strong Coupling}\label{sec:limiting}
%
We construct singular perturbation expansions to analyze the limiting behaviors of Eq.~\eqref{eq:eig_eq} when the interlayer-coupling strength $\omega$ is very small (i.e., layer decoupling) or very large (i.e., layer aggregation). These results provide insights into our supracentrality framework and can aid in the selection of appropriate parameter values.

\subsection{Layer Decoupling in the Weak-Coupling Limit}\label{sec:weak}
%
\new{We first study supracentralities in the $\omega\to 0^+$ limit.
We did not study this limit in our previous work \cite{taylor2017eigenvector} on temporal centralities.
We analyze how the eigenvalues and eigenvectors of $\mathbb{C}(\omega)$ for small $\omega$ are determined by the eigenvalues and eigenvectors of the individual layers' centrality matrices ${\bf C}^{(t)}$.}


\new{
\begin{lemma}[Layer Decoupling   at $\omega=0$] \label{thm:uncoupled}
Let $\mu_i^{(t)}$, ${\bf v}^{(i,t)}$  and ${\bf u}^{(i,t)}$  denote the eigenvalues and corresponding right and left eigenvectors of {the} $N\times N$ centrality matrices ${\bf C}^{(t)}$ for $t\in\{1,\dots,T\}$. It follows that each $\mu_i^{(t)}$ is an eigenvalue of the associated supracentrality matrix $\mathbb{C}(0)=\mathrm{diag}[{\bf C}^{(1)}$,\dots,${\bf C}^{(T)}]$ with corresponding left eigenvector $\mathbbm{u}^{(i,t)} = {\bf e}^{(t)} \otimes  {\bf u}^{(i,t)}$ and right eigenvector $\mathbbm{v}^{(i,t)} = {\bf e}^{(t)} \otimes  {\bf v}^{(i,t)}$, where ${\bf e}^{(t)}$ denotes a length-$T$ unit vector that consists of zeros in all entries except for the $t$-th entry (which is a $1$). 

\begin{proof}
See Appendix~\ref{app:uncoupled}.
\end{proof}

\end{lemma}
}


\new{
\begin{remark}\label{remark:ahdddh}
Each of the vectors $\mathbbm{u}^{(i,t)} $ and $ \mathbbm{v}^{(i,t)} $ 
consists of $T$ blocks, and each block is a length-$N$ vector that consists of zeros except for the $t$-th block, which is the associated
left or right eigenvector of ${\bf C}^{(t)}$.
\end{remark}
}


\new{
\begin{remark}\label{remark:ahh}
If there exist multiple eigenvalues $\mu_i^{(t)}$ of ${\bf C}^{(t)}$ that are equal (that is, $\mu_i^{(t)}  =\lambda$ for any $(i,t)$ in some set $\mathcal{P}$), then any linear combination of their associated eigenvectors is also an eigenvector:
\begin{align}\label{eq:linear_combo}
\mathbb{C}(0) \left( \sum_{(i,t)\in\mathcal{P}}  \alpha_{i,t} \mathbbm{v}^{(i,t)} \right)
= \sum_{(i,t)\in\mathcal{P}}  \alpha_{i,t}  \mathbb{C}(0) \mathbbm{v}^{(i,t)} 
&= \sum_{(i,t)\in\mathcal{P}}  \alpha_{i,t}  \lambda \mathbbm{v}^{(i,t)}\nonumber\\
&= \lambda \left( \sum_{(i,t)\in\mathcal{P}}  \alpha_{i,t} \mathbbm{v}^{(i,t)} \right)  \,.
\end{align}
One can show a similar result for the left eigenvectors. Consequently, the eigenspace that is associated with each eigenvalue is an invariant subspace.
\end{remark}

We now turn our attention to the eigenspace of the dominant eigenvalue (i.e., the dominant eigenspace). Let $\lambda_{\rm{max}}(0)$ denote the largest positive (i.e., dominant)  eigenvalue of $\mathbb{C}(0)$, and let $\mathcal{P} = \{t : \mu_1^{(t)} = \lambda_{\rm{max}}(0) \}$ denote the set of indices for layers whose  largest positive eigenvalue $\mu_1^{(t)} $ is equal to
$ \lambda_{\rm{max}}(0)$. We assume that each layer's dominant eigenvalue $\mu_1^{(t)}$ is simple (i.e, $ \mu_j^{(t)}<\mu_1^{(t)}$ for any $j\ge 2$) and that its right and left eigenvectors are unique. This implies that $i=j=1$ for our notation in Eq.~\eqref{eq:linear_combo}, so we 
can consider a set \new{$\mathcal{T}$} of layers rather than a set $\mathcal{P}$ of node-layer pairs.}

We now present a key result for the  dominant eigenvectors of 
\new{$\lim_{\omega \to 0^+}\mathbb{C}(\omega) $.}


\begin{theorem}[Weak-Coupling Limit of Dominant Eigenvectors] \label{thm:uncoupled2}
Let    $\mathbbm{v}^{(1)}(\omega)$ and $\mathbbm{u}^{(1)}(\omega)$, respectively, be the  right and left dominant eigenvectors of a supracentrality matrix $\mathbb{C}(\omega)$ under the assumptions of Thm.~\ref{thm:unique}.
It then follows that the  $\omega \to 0^+$ limits of $\mathbbm{u}^{(1)}(\omega)$ and $\mathbbm{v}^{(1)}(\omega)$ are 
\begin{equation}\label{eq:lim_uv}
	\mathbbm{v}^{(1)}(\omega)  \to \sum_{t\in\new{\mathcal{T}}}   \alpha_{t} \mathbbm{v}^{(1,t)}\,, ~~
\mathbbm{u}^{(1)}(\omega)  \to \sum_{t\in\new{\mathcal{T}}}  \beta_{t} \mathbbm{u}^{(1,t)} \, ,
\end{equation}
where the vectors $\bm{\alpha} = [\alpha_{1},\dots,\alpha_{T}]^\new{*}$ and $\bm{\beta}=[{\beta}_{1},\dots,\beta_{T}]^\new{*}$, which have nonnegative entries that satisfy $\sum_t \alpha_t^2 =\sum_t \beta_t^2 =1$, are  positive solutions to \new{the dominant-eigenvalue equations}
\begin{equation}\label{eq:baah}
	\bm{X} \bm{ \alpha} = \lambda_{1}  \bm{ \alpha}\,, ~~
 {\bm{X}}^\new{*} \bm{ \beta} = \lambda_{1}  \bm{ \beta}\,,
\end{equation}
where $\lambda_{1}$ is an eigenvalue that needs to be determined, the entries of $\bm{X}$ are
\begin{align}\label{eq:Xtt}
	{X}_{tt'} &= \tilde{{A}}_{tt'}  \frac{ \langle {\bf u}^{(1,t)} , {\bf v}^{(1,t')} \rangle}{\langle{\bf u}^{(1,t)},   {\bf v}^{(1,t)} \rangle }   \chi(t)\chi(t') \,,
\end{align}
and $\chi(t)=\sum_{t'\in\new{\mathcal{T}}} \delta_{tt'}$ is an indicator function (with $\chi(t)=1$ if $t\in\new{\mathcal{T}}$ and $\chi(t)=0$ otherwise). \new{(Recall that $*$ is our notation for transpose operator.)}

\begin{proof}
See Appendix~\ref{app:uncoupled2}.
\end{proof}
\end{theorem}

\new{
\begin{remark}\label{remark:singlet_0}
We obtained Eq.~\eqref{eq:Xtt} using singular perturbation theory in the limit $\omega\to0^+$. One can understand why the limit is singular by considering the dimension $f(\omega)=\mathrm{dim}(\mathrm{null}(\mathbb{C}(\omega)-\lambda_{\rm{max}}(\omega)\mathbb{I}))$ of the eigenspace that is associated with the dominant eigenvalue $ \lambda_{\rm{max}}(\omega)$ of $\mathbb{C}(\omega)$. For any $\omega>0$, Thm.~\ref{thm:unique} (i.e., our Perron--Frobenius theorem) guarantees that $f(\omega)=1$. 
However, Eq.~\eqref{eq:linear_combo} implies that when $\omega=0$, there are $|\new{\mathcal{T}}|$ eigenvectors that are associated with eigenvalue $ \lambda_{\rm{max}}(0)$, which in turn implies that $f(0)=|\new{\mathcal{T}}|$. If $|\new{\mathcal{T}}|>1$, then $\lim_{\omega\to 0^+} f(\omega)\not =  f(0)$ and $\omega=0$ is a singular point   of the dominant eigenspace.
\end{remark}
}


We now present three corollaries that consider Thm.~\ref{thm:uncoupled2} under various restrictions on the centrality matrices. We first consider the limiting behavior when the layers' centrality matrices all have the same spectral radius, as is the case for PageRank matrices (because a PageRank matrix is a transition matrix of a Markov chain) or if one rescales the centrality matrices to have the same spectral radius.

\begin{corollary}[Weak-Coupling Limit for 
Centrality Matrices with the Same Spectral Radius] \label{corr:equivalent}
Under the assumptions of Thm.~\ref{thm:uncoupled2} and also assuming that all centrality matrices have the same spectral radius (i.e., $\lambda_{\rm{max}} = \mu_1^{(t)}$ for all $t$), it follows that $\new{\mathcal{T}}=\{1,\dots,T\}$ and $\chi(t)=1$. Additionally, Eq.~\eqref{eq:Xtt} takes the form
\begin{align}\label{eq:Xtt2}
	{X}_{tt'}  &= \tilde{{A}}_{tt'}  \frac{ \langle {\bf u}^{(1,t)} , {\bf v}^{(1,t')} \rangle}{\langle{\bf u}^{(1,t)},    {\bf v}^{(1,t)} \rangle }  \,.
\end{align}
\end{corollary}

We next consider when the layers' centrality matrices are symmetric, which is the case for hub/authority matrices and for symmetric adjacency matrices.

\begin{corollary}[Weak-Coupling Limit for Symmetric Centrality Matrices] \label{corr:symm} \\
Under the assumptions of Thm.~\ref{thm:uncoupled2} and also assuming that all centrality matrices are symmetric, ${\bf u}^{(1,t)} = {\bf v}^{(1,t')} $ and Eq.~\eqref{eq:Xtt} takes the form
\begin{align}\label{eq:Xtt3}
	{X}_{tt'}  &= \tilde{{A}}_{tt'}   \chi(t)\chi(t') \,.
\end{align}
\end{corollary}

When the centrality matrix of a single layer has the largest spectral radius, which one often expects to occur for adjacency matrices and hub/authority matrices (unless the network layers have symmetries that yield repeated spectral radii across layers), the limiting behavior of the eigenvector is that it localizes onto a single dominating layer.

\begin{corollary}[Weak-Coupling-Induced Eigenvector Localization onto a Dominating Layer] \label{corr:dominate}
Under the assumptions of Thm.~\ref{thm:uncoupled2} and also assuming that one layer has a spectral radius that is larger than all others (i.e., $\lambda_{\rm{max}} = \mu_1^{(t)}$ for a single layer $t=\new{\tau}$), then as $\omega\to0^+$, it follows that
\begin{align}
	\mathbbm{v}(\omega)  \to  \mathbbm{v}^{(1,\new{\tau})}\,, ~~ 
	\mathbbm{u}(\omega)  \to  \mathbbm{u}^{(1,\new{\tau})} \,.  
\end{align}
\end{corollary}

Understanding whether or not the dominant eigenvector localizes onto a single layer, onto several layers (i.e., as given by the function $\chi(t)$), or does not localize has significant practical consequences. In some situations, it can be appropriate to allow eigenvector localization \cite{pastor2018eigenvector}, whereas it can be beneficial to avoid localization in others \cite{martin2014localization}. 
\new{Lemma~\ref{thm:uncoupled} and Corollaries \ref{corr:symm}--\ref{corr:dominate}} characterize localization in the weak-coupling limit and are useful for practitioners to make informed
choices about which centrality matrices to use.

\subsection{Layer Aggregation in the Strong-Coupling Limit}\label{sec:strong}
%
We study Eq.~\eqref{eq:eig_eq} in the limit as $\omega\to \infty$ (or, equivalently, as $\epsilon := \omega^{-1}\to 0^+$). The results of this subsection generalize those of \cite{taylor2017eigenvector}, where we assumed that $\tilde{\bm{ A}}$ encodes adjacent-in-time coupling. In the present discussion, by contrast, we allow $\tilde{\bm{ A}}$ to be from a much more general class of matrices, including asymmetric matrices that encode directed interlayer couplings.

Consider the scaled supracentrality matrix 
\begin{equation}\label{eq:newC}
	\tilde{\mathbb{C}}(\epsilon)  = \epsilon \mathbb{C}(\epsilon^{-1}) = \epsilon\hat{\mathbb{C}} +   \hat{\mathbb{A}}\,, 
\end{equation}
which has  eigenvectors $\tilde{\mathbbm{u}}(\epsilon)$ and $\tilde{\mathbbm{v}}(\epsilon)$ that are identical to those of $\mathbb{C}(\omega) $ 
(specifically, $\tilde{\mathbbm{u}}(\epsilon)={\mathbbm{u}}(\epsilon^{-1})$ and $\tilde{\mathbbm{v}}(\epsilon)={\mathbbm{v}}(\epsilon^{-1})$). Its eigenvalues $\{ \tilde{\lambda}_i\}$ are scaled by $\epsilon$; specifically, $ \tilde{\lambda}_i(\epsilon) = \epsilon\lambda_i(\epsilon^{-1})$. 

To facilitate our presentation, we define a permutation operator for $NT \times NT$ matrices.

\begin{definition}[Node-Layer-Reordering Stride Permutation]\label{def:stride}
The matrix\, $\mathbb{P}$ is a \emph{$T$-stride permutation matrix} of size $NT\times NT$ if it has entries that take the form \cite{golub} 
\begin{equation}\label{eq:stride}
	[\mathbb{P}]_{kl} = \left\{ \begin{array}{rl}
1\,,& l=\lceil k/N\rceil+T\,[(k-1)\bmod N] \\
 0\,,& \textrm{otherwise} \,. 
\end{array} \right.
\end{equation}
Therefore, $ \left( \tilde{\bm{A}}  \otimes  \mathbf{I} \right)  = \mathbb{P}  \left( \mathbf{I} \otimes \tilde{\bm{A}}  \right)\mathbb{P}^\new{*}$.
\end{definition}

\begin{remark}
The stride-permutation matrix is unitary, and it simply changes the ordering of node-layer pairs. Before the stride permutation (which is a type of graph isomorphism \cite{mikko-tnse}), a supracentrality matrix has entries that are ordered first by node $i$ and then by layer $t$ (i.e., $(i,t) = (1,1), (2,1), (3,1), \dots$). After the stride permutation, the entries are ordered first by layer $t$ and then by node $i$ (i.e., $(i,t) = (1,1), (1,2), (1,3), \dots$). 
\end{remark}

We now present our main findings for the strong-coupling regime.

\begin{lemma}[Singularity at Infinite Coupling] \label{thm:singularity}
Let $\tilde{\mu}_t$ denote the eigenvalues of $\tilde{\bm{A}}$, and let $\tilde{\bm{v}}^{(t)}$ and $\tilde{\bm{u}}^{(t)}$, respectively, be the corresponding right and left eigenvectors. We assume that the eigenvalues are simple, and we order them such that $\tilde{\mu}_1$ is the largest eigenvalue.  We also let $\mathbb{P}$ denote the stride permutation matrix from Eq.~\eqref{eq:stride}.

\new{
For $\epsilon=0$, each $\tilde{\mu}_t$ is an eigenvalue of $\mathbb{C}(\epsilon)$}
given by Eq.~\eqref{eq:newC}, \new{and the associated}
$N$-dimensional 
right and left eigenspaces are spanned by the eigenvectors \new{$\mathbb{P}\tilde{\mathbbm{v}}^{(t)}$ and $\mathbb{P}\tilde{\mathbbm{u}}^{(t)}$}, respectively, with the general form
\begin{equation}\label{eq:general_evec}
	\tilde{\mathbbm{v}}^{(t)} = \sum_j \tilde{\alpha}_{tj} \mathbb{P}\tilde{\mathbbm{v}}^{(t,j)}\,, ~ ~
	\tilde{\mathbbm{u}}^{(t)} = \sum_j \tilde{\beta}_{tj} \mathbb{P}\tilde{\mathbbm{u}}^{(t,j)}  \,,
\end{equation}
where the constants $\tilde{\alpha}_{tj}$  and $\tilde{\beta}_{tj}$ must satisfy $\sum_j\tilde{\alpha}_{tj}^2=\sum_j\tilde{\beta}_{tj}^2=1$ to ensure that $\| \tilde{\mathbbm{u}}^{(t)}\|_2 = \| \tilde{\mathbbm{v}}^{(t)}\|_2=1$.
The associated length-$NT$ vectors are $\tilde{\mathbbm{v}}^{(t,j)} = \tilde{{\bf e}}^{(j)} \otimes  {\bf v}^{(t)}$ and $\tilde{\mathbbm{u}}^{(t,j)} = \tilde{{\bf e}}^{(j)} \otimes  {\bf u}^{(t)} $,
where $\tilde{{\bf e}}^{(j)}$ is a length-$N$ unit vector that consists of zeros in all entries except for entry $j$, which is $1$. Therefore,  $\tilde{\mathbbm{u}}^{(t,j)}$ ({respectively,} $\tilde{\mathbbm{v}}^{(t,j)}$) consists of zeros, except in the $j$-th block of size $T$, which consists of a left (respectively, right) eigenvector of $\tilde{\bm{A}}$.  

\begin{proof}
See Appendix~\ref{app:singularity}.
\end{proof}
\end{lemma}

\begin{remark}
It is straightforward to also obtain the general form of eigenvectors for eigenvalues $\{\mu_t\}$ whose multiplicity is larger than $1$. For example, if the eigenvalue $\tilde{\mu}_t$ of $\tilde{\bm{A}}$ has multiplicity $q$, then $\tilde{\lambda}_i(0)=\tilde{\mu}_t$ has multiplicity $qN$ for the matrix $\mathbb{C}(0)$. However, the notation becomes slightly more cumbersome, and we will not study such cases in this paper.  
\end{remark}

\begin{theorem}[Strong-Coupling Limit of Dominant Eigenvectors] \label{thm:singular_limit}
Let $\tilde{\mu}_1$ denote the dominant eigenvalue (which we assume to be simple) of the interlayer-adjacency matrix $\tilde{\bm{A}}$, and let $\tilde{\bm{v}}^{(1)}$ and $\tilde{\bm{u}}^{(1)}$ be its associated right and left eigenvectors. We assume that the constraints of Thm.~\ref{thm:unique} are satisfied, such that the supracentrality matrix $\mathbb{C}(\epsilon)$ given by Eq.~\eqref{eq:supracentrality} is nonnegative, irreducible, and aperiodic. It then follows that the  {largest positive} eigenvalue $\tilde{\lambda}_{\rm{max}}(\epsilon)$ and its associated eigenvectors, $\mathbbm{u}^{(1)}(\epsilon)$ and $\mathbbm{v}^{(1)}(\epsilon)$, of $\mathbb{C}(\epsilon)$ converge as $\epsilon \to 0^+$ as follows:
\begin{equation}
	\tilde{\lambda}_{\rm{max}}(\epsilon)     \to\tilde{ \mu}_1   \,,~~ 
	\tilde{\mathbbm{v}}^{(1)}(\epsilon) \to \sum_j \tilde{\alpha}_{j} \mathbb{P}\tilde{\mathbbm{v}}^{(1,j)} \,,~~
	\tilde{\mathbbm{u}}^{(1)}(\epsilon) \to \sum_j \tilde{\beta}_{j} \mathbb{P}\tilde{\mathbbm{u}}^{(1,j)}\, , 	\label{eq:conver2}
\end{equation}
where $\mathbb{P}$ is the stride permutation from Eq.~\eqref{eq:stride}, the vectors $\tilde{\mathbbm{u}}^{(1,j)}$ and $\tilde{\mathbbm{v}}^{(1,j)}$ are defined in Thm.~\ref{thm:singularity}, and the constants $\tilde{\beta}_i$  and $\tilde{\alpha}_i$ solve the  dominant-eigenvalue equations 
\begin{equation}\label{eq:alpha_beta}
	\tilde{\mathbf{X}}  \tilde{\bm{\alpha}} =  \tilde{\mu}_1 \tilde{\bm{\alpha}}\,, ~~ 
	\tilde{\mathbf{X}}^\new{*}  \tilde{\bm{\beta}} =  \tilde{\mu}_1    \tilde{\bm{\beta}} \,,
\end{equation}
where 
\begin{align}
	\tilde{X}_{ij} 	
		&=   \sum_{t} {C}^{(t)}_{ij} \frac{\tilde{u}_t^{(1)} \tilde{v}_t^{(1)}}{\langle \tilde{\bm{u}}^{(1)}, \tilde{\bm{v}}^{(1)} \rangle} \,. \label{eq:M1star_0_q} 
\end{align} 
\begin{proof}
See Appendix~\ref{app:singular_limit}. 
\end{proof}
\end{theorem}

Equation~\eqref{eq:M1star_0_q} indicates that the strong-coupling limit effectively aggregates the centrality matrices $\{ {\bf C}^{(t)} \}$ across time via a weighted average, with weights that depend on the right and left dominant eigenvectors of the interlayer-adjacency matrix $\tilde{\bm{A}}$. This result generalizes Eq.~(4.13) of Taylor et al.~\cite{taylor2017eigenvector}, who assumed that $\tilde{\bm{A}}$ is symmetric (so that $\tilde{u}_t^{(1)}= \tilde{v}_t^{(1)}$). We recover the result in \cite{taylor2017eigenvector} with the following corollary.

\begin{corollary}[Strong-Coupling Limit of Eigenvector-Based Centralities for Multilayer Networks with Adjacent-in-Time, Uniform, and Diagonal Coupling \cite{taylor2017eigenvector}]\label{corr:nn}
For undirected, adjacent-in-time interlayer coupling (i.e., $\tilde{A}_{tt'}=1$ for $|t-t'|=1$ and $\tilde{A}_{tt'}=0$ otherwise), the $\epsilon\to0^+$ limit of the largest eigenvalue is $\tilde{\lambda}_1(\epsilon) \to
2 \cos\left({\frac{\pi}{T+1}}\right) $. When $\epsilon\to0^+$, the  right and left dominant eigenvectors satisfy Eqs.~\eqref{eq:conver2}--\eqref{eq:alpha_beta}, with
\begin{equation*}
	\tilde{\mathbf{X}} =   \sum_{t} \mathbf{C}^{(t)}  \frac{\sin^2\left( \frac{\pi t}{T+1}\right)}{\sum_{t=1}^T \sin^2\left( \frac{\pi t}{T+1}\right)}\,.
\end{equation*}	
\end{corollary}

\begin{corollary}[Strong-Coupling Limit of Eigenvector-Based Centralities for Multilayer Networks with All-to-All, Uniform, and Diagonal Coupling]\label{corr:aa}
For all-to-all coupling (with self-edges), $\tilde{\bm{A}} = \bm{1}\bm{1}^\new{*} $; the $\epsilon\to0^+$ limit of the largest eigenvalue is 
$\lambda_{\rm{max}}(\epsilon) \to \tilde{\mu}_1=N$; and, when $\epsilon\to0^+$, the  right and left dominant eigenvectors satisfy Eqs.~\eqref{eq:conver2}--\eqref{eq:alpha_beta}, with $\tilde{\mathbf{X}} =  T^{-1}\sum_{t} \mathbf{C}^{(t)} $.

\begin{proof}
In this case, the largest eigenvalue of $\tilde{\bm{A}}$ is $\mu_1 = N $, and the right and left dominant eigenvectors have the same value in each component, with entries $u_t^{(1)}= v_t^{(1)}=T^{-1/2}$.
\end{proof}
\end{corollary}


\begin{corollary}[Strong-Coupling Limit of of Eigenvector-Based Centralities for Multilayer Networks with Rank-$1$, Uniform, and Diagonal Coupling]\label{corr:rank1}
For a rank-$1$ interlayer-adjacency matrix,  $\tilde{\bm{A}} =   \tilde{\bm{w} }\tilde{\bm{w} }^\new{*} $, the $\epsilon\to0^+$ limit of the largest eigenvalue is  $\lambda_{\rm{max}}(\epsilon) \to 1$; and, when $\epsilon\to0^+$, the  right and left dominant eigenvectors satisfy Eqs.~\eqref{eq:conver2}--\eqref{eq:alpha_beta}, with
$\tilde{\mathbf{X}} = \sum_{t} \mathbf{C}^{(t)}\tilde{ w}_t^2$.
\begin{proof}
This follows from the fact that the largest eigenvalue of $\tilde{\bm{A}}$ is $\tilde{\mu}_1 = 1$, with associated eigenvectors $\tilde{u}_t^{(1)}= \tilde{v}_t^{(1)}=\tilde{w}_t$.
\end{proof}
\end{corollary}


\section{Case Studies with Empirical Data}\label{sec:data}
%

We now apply our supracentrality framework to study networks that we construct from two data sets. In Sec.~\ref{sec:mux}, we examine a multiplex network that encodes flight patterns between European airports, with layers representing different airline companies \cite{cardillo2013emergence}. In Sec.~\ref{sec:temp}, we examine a temporal network that encodes the exchange of mathematicians between mathematical-sciences doctoral programs in the United States \cite{taylor2017eigenvector}. These experiments explore different interlayer-coupling strengths and topologies, and they also illustrate strategies for how to flexibly apply our supracentrality framework to diverse applications.

\begin{figure}[b!]
\centering
\includegraphics[width=\linewidth]{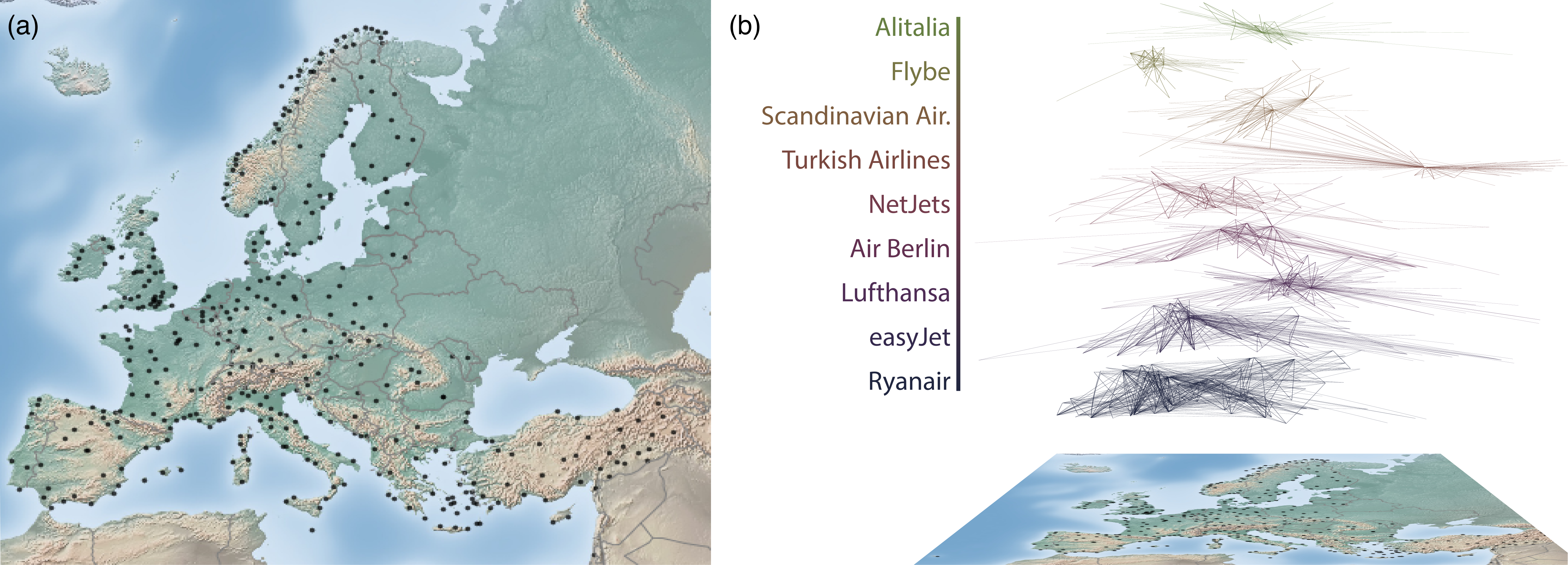}
\vspace{-0.3cm}
\caption{
Multilplex network that encodes flights between European airports.
(a)~A map of airport locations.
(b)~An illustration of the nine network layers (and hence airlines) that have the most edges.  {(We created this map using the Python module {\sc Basemap} \cite{basemap}.)}
}\label{fig:mux_nice}
\end{figure}

\begin{table}[t!]
\caption{
A list of European airline companies, which we represent as layers in a multiplex  network. For each layer, we report 
  the number $M_l$ of undirected edges and the spectral radius $\lambda_1^{(t)}$ of its associated adjacency matrix ${\bf A}^{(t)}$. We have chosen the ordering to match the one in \cite{cardillo2013emergence}.
}
\centering{
\scriptsize{
\begin{tabular}{cc}
\begin{tabular}{cccc}
\hline\hline 
Layer $t$ & Airline Name & $M_t$ & $\lambda_1^{(t)}$ \\  [0.5ex] 
\hline  
1  & Lufthansa &244 &14.5\\
2  & Ryanair &601 &19.3\\
3  & easyJet&307&14.0\\
4  & British Airways &66&6.6\\
5  & Turkish Airlines&118&9.9\\
6  & Air Berlin&184&11.3\\
7  & Air France&69&7.2\\
8  & Scandinavian Air. &110&8.9\\
9  & KLM &62&7.9\\
10 & Alitalia &93&8.8\\
11 & Swiss Int. Air Lines &60&7.3\\
12 & Iberia &35&5.8\\
13 & Norwegian Air Shu. &67&8.1\\
14 & Austrian Airlines &74&8.1\\
15 & Flybe &99&8.5\\
16 & Wizz Air &92&6.5\\
17 & TAP Portugal&53&7.0 \\
18 & Brussels Airlines &43&6.6\\
19 & Finnair &42 &6.4\\
[1ex]  
\hline \hline
\end{tabular}&
\begin{tabular}{cccc}
\hline\hline 
Layer $t$ & Airline Name & $M_t$ & $\lambda_1^{(t)}$ \\  [0.5ex] 
\hline  
20 & LOT Polish Air.  &55&6.8\\
21 & Vueling Airlines &63&6.8\\
22 & Air Nostrum &69&6.4\\
23 & Air Lingus &108&6.7\\
24 & Germanwings &67&7.4\\
25 & Pegasus Airlines &58 &6.7\\
26 & NetJets &180 &8.2\\
27 & Transavia Holland &57&6.0\\
28 & Niki &37 &4.7\\
29 & SunExpress &67 &7.8\\
30 & Aegean Airlines &53 &6.5\\
31 & Czech Airlines &41 &6.4\\
32 &  European Air Trans. &73 &6.8\\
33 &  Malev Hungarian Air. &34 &5.8\\
34 &  Air Baltic &45 &6.4\\
35 &  Wideroe &40 &5.6\\
36 &  TNT Airways &61&6.2\\
37 & Olympic Air &43 &6.2 \\
~\\
[1ex]  
\hline \hline
\end{tabular}
\end{tabular}}
\label{table:Airlines} }
\end{table}

\subsection{Rankings of European Airports in a Multiplex Airline Network \cite{cardillo2013emergence}}\label{sec:mux}
%
We first apply our supracentrality framework to study the importances of European airports using an empirical multiplex network of the flight patterns for $T=37$ airline companies \cite{cardillo2013emergence}. See Fig.~\ref{fig:mux_nice} for a map of the airports and a visualization of the network layers that have the most edges. The network's nodes represent European airports; we consider only the $N=417$ nodes in the largest connected component of the network that is associated with the sum of the layers' adjacency matrices. We couple the layers using all-to-all coupling (so $\tilde{\bm{A}} = \bm{1}\bm{1}^\new{*}$), and we calculate eigenvector supracentralities with the centrality matrices from Definition \ref{def:Eigen}. See Table~\ref{table:Airlines} for a list of the airline companies that are associated with the network layers. For each layer, we indicate the number $M_t=\frac{1}{2}\sum_{ij} A^{(t)}_{ij}$ of intralayer edges and the spectral radius $\lambda_1^{(t)}$ of the layer's associated adjacency matrix ${\bf A}^{(t)}$.

\begin{figure}[h!]
\centering
\includegraphics[width=.95\linewidth]{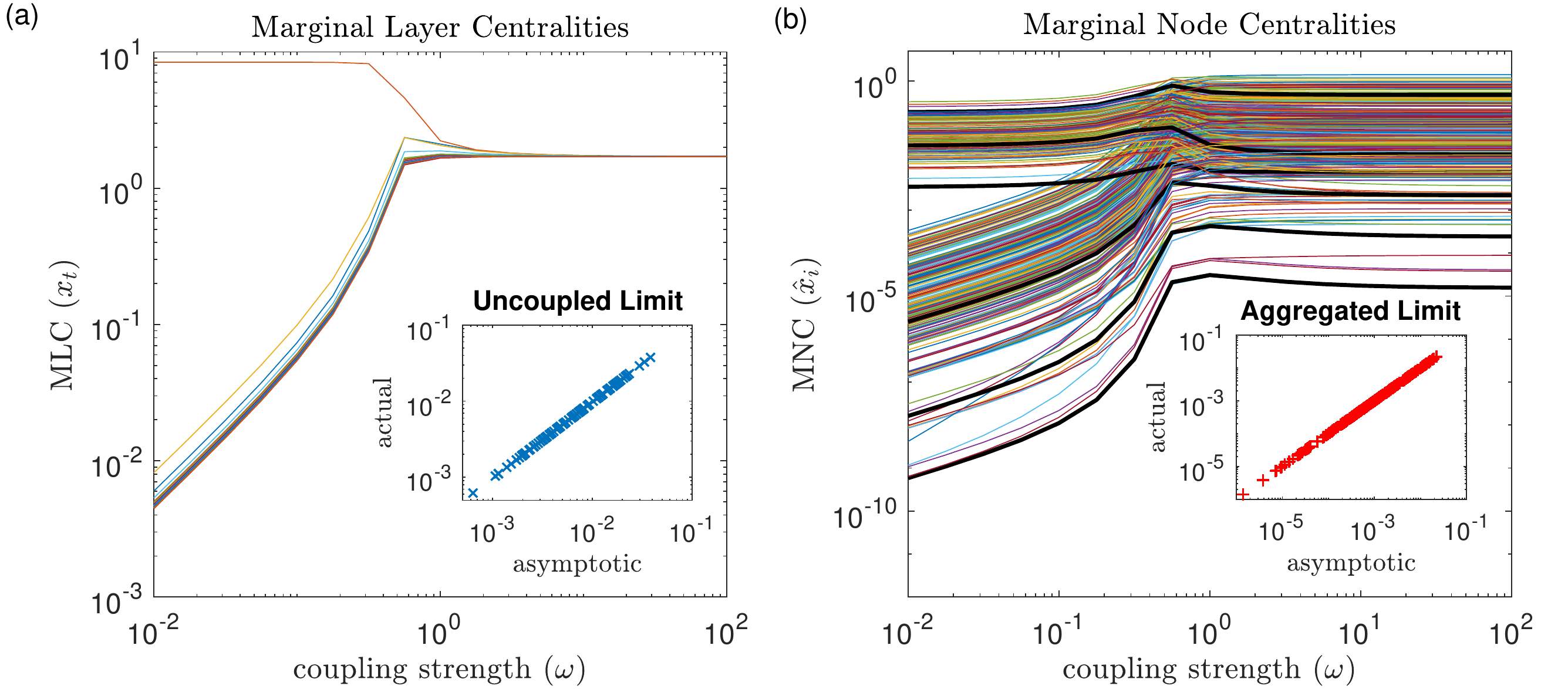}
\vspace{-.2cm}
\caption{
Marginal layer centralities (MLCs) and marginal node centralities (MNCs) for a multiplex European airline transportation network of the flight patterns of 37 airlines 
\cite{cardillo2013emergence}. We couple the layers with all-to-all coupling and examine interlayer-coupling strengths of $\omega\in[10^{-2},10^{2}]$. The insets in panels (a) and (b) compare the calculated conditional node centralities for $\omega=10^{-2}$ and $\omega=10^{2}$, respectively, to the asymptotic values from Thms.~\ref{thm:uncoupled2} and \ref{thm:singular_limit}.
}\label{fig:mux_marg_eig}
\end{figure}

For each airport, we compute the joint, marginal, and conditional centralities for a range of interlayer-coupling strengths $\omega$. In Fig.~\ref{fig:mux_marg_eig}, we plot the MLCs and MNCs (see Definition~\ref{def:marg}). In Fig.~\ref{fig:mux_marg_eig}(a), we see that for large $\omega$, all layers have similar importances; by contrast, for small $\omega$, one layer is much more important because the eigenvector-localization phenomenon that we described in Cor.~\ref{corr:dominate}. Specifically, the layer that represents Ryanair dominates for small $\omega$, as its adjacency matrix has the largest spectral radius. (It also has the most edges.)  A previous investigation of multilayer centralities in this data set \cite{rahmede2017centralities} also identified 
Ryanair as the most important airline.

The insets in Fig.~\ref{fig:mux_marg_eig} give the calculated conditional node centralities for $\omega=10^{-2}$ (in panel (a)) and $\omega=10^{2}$ (in panel (b)) versus the asymptotic values in the associated limits for $\omega\to0^+$ (see Thm.~\ref{thm:uncoupled2}) and $\omega\to\infty$ (see Thm.~\ref{thm:singular_limit}), demonstrating that they are in excellent agreement. One can also observe in Fig.~\ref{fig:mux_marg_eig}(b) that there is not a simple transition between these two limits. Specifically, the thick black curves highlight a few airports whose MNCs have a peak for intermediate values of $\omega$. 
These airports are more important if one considers the airline network as a multiplex network than if one considers the layers in isolation or in aggregate.

\begin{table}[h]
\caption{European airports with the largest MNCs for interlayer-coupling strengths $\omega$ in the regimes of weak ($\omega=0.01$), intermediate ($\omega=1$), and strong ($\omega=100$) coupling. We give the results of our computations of eigenvector supracentralities with all-to-all, uniform, and diagonal coupling between layers. We identify each airport by its International Civil Aviation Organization (ICAO) code. 
}
\vspace{-.1cm}
\centering{\small{
\begin{tabular}{cccc} 
\begin{tabular}{c}
~\\
\hline\hline 
Rank \\
[0.5ex] 
\hline  
1 \\
2 \\
3 \\
4 \\
5 \\
6 \\
7 \\
8 \\
9 \\
10 \\
[1ex]  
\hline\hline
\end{tabular}&
\begin{tabular}{c}
$\omega=0.01$\\
\begin{tabular}{cc}
\hline\hline 
Airport  & MNC\\  [0.5ex] 
\hline  
EGSS & 0.329 \\  
EIDW  & 0.286\\
LIME &  0.254\\
EBCI & 0.201\\
LEMD &0.193 \\
LEAL  &0.190 \\
EDFH & 0.189\\
LIRA & 0.184\\
LEGE & 0.176\\
LEPA & 0.166\\
[1ex]  
\hline \hline
\end{tabular}
\end{tabular}&
\begin{tabular}{c}
$\omega=1$\\
\begin{tabular}{cc}
\hline\hline 
Airport & MNC  \\  [0.5ex] 
\hline  
LEMD & 1.379\\
EHAM & 1.296\\
LEBL & 1.257\\
EDDM & 1.171\\
LIRF & 1.150\\
EDDF & 1.121\\
EDDL & 1.105\\
LFPG & 1.091\\
LOWW & 1.066\\
LIMC & 0.968\\
[1ex]  
\hline \hline
\end{tabular}
\end{tabular}&
\begin{tabular}{c}
$\omega=100$\\
\begin{tabular}{cc}
\hline\hline 
Airport  & MNC  \\  [0.5ex] 
\hline  
EHAM & 1.406\\
LEMD & 1.400\\
LIRF & 1.206\\
LOWW & 1.198\\
LEBL & 1.193\\
EDDM & 1.160\\
LFPG & 1.157\\
EDDF & 1.134\\
EDDL & 1.128\\
LSZH & 1.017\\
[1ex]  
\hline \hline
\end{tabular}
\end{tabular}
\end{tabular}}
\label{table:tops} 
}
\end{table}

In Table~\ref{table:tops}, we list the airports with the largest MLCs for eigenvector supracentrality for small, intermediate, and large values of $\omega$. As expected, for large and small $\omega$, the top airports correspond to the top airports (i.e., those with the largest eigenvector centralities) that are associated with the aggregation of layers and the Ryanair network layer, respectively. The top-ranked airports for $\omega=1$ have a large overlap with those for $\omega=100$. The top airport, Adolfo Su\'arez Madrid--Barajas Airport (LEMD), is particularly interesting. LEMD is the only airport that ranks in the top 10 for both the Ryanair layer and the layer-aggregated network; this contributes to its having the top rank for this intermediate value of $\omega$. We highlight similar ranking boosts for other airports with solid black curves in Fig.~\ref{fig:mux_marg_eig}(b). We also note that LEMD was identified in previous investigations \cite{iacovacci2016functional,tudisco2018node} as one of the most important airports in this data set.

\begin{figure}[t]
\centering
\includegraphics[width=.85\linewidth]{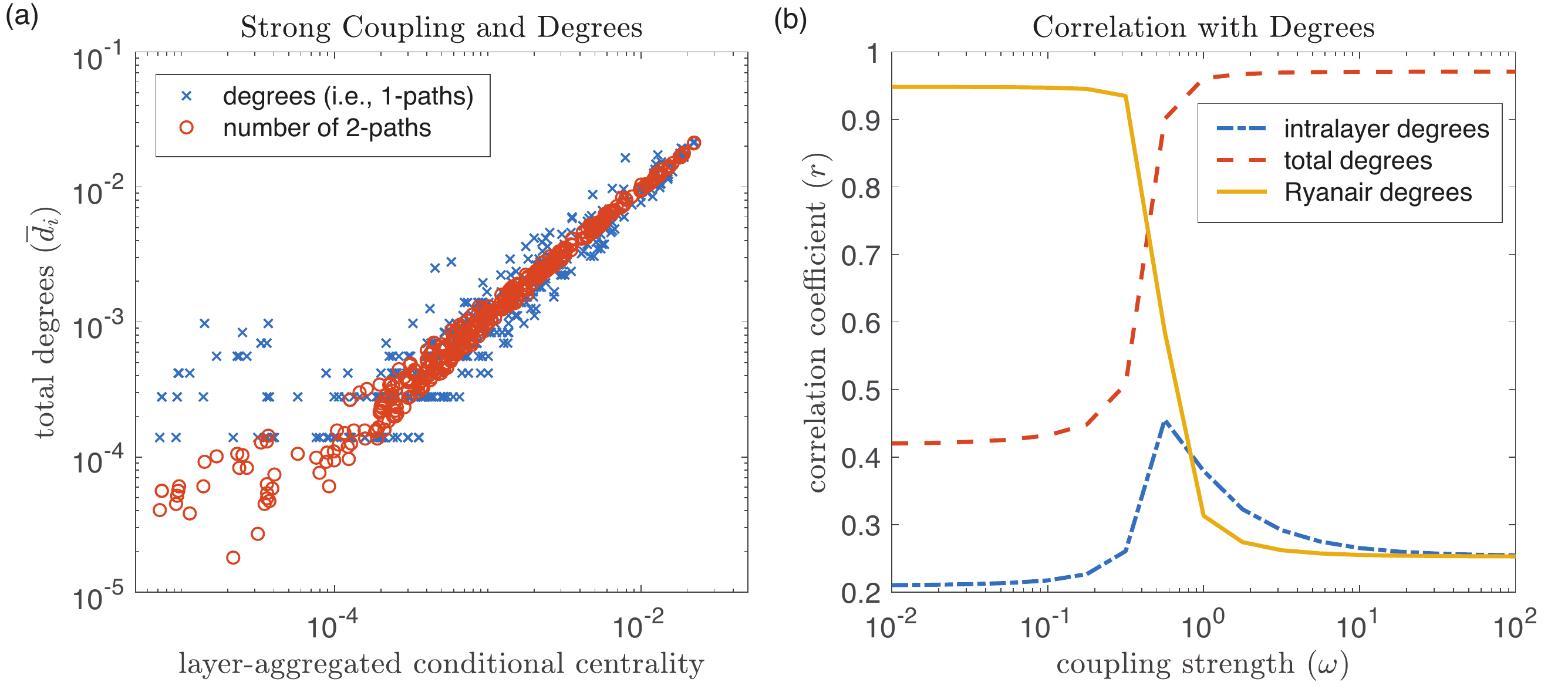}
\vspace{-.2cm}
\caption{
Eigenvector supracentralities for the multiplex European airline network.
(a)~For $\omega=100$, the airports' MNCs correlate strongly with the layer-aggregated degrees $d_i  = \sum_{t,j} A_{ij}^{(t)}$ and with the total number of length-2 paths (summed across layers) that emanate from each node. 
To facilitate this comparison, we normalize the vectors. In the legend, we write ``$k$-path'' as shorthand terminology for ``length-$k$ path''.
(b)~We compute the Pearson correlation coefficients to compare the airports' eigenvector supracentralities to three different notions of node degree that one can define for a multiplex network:  
(dot-dashed blue curve) intralayer degrees $d_i^{(t)} = \sum_j A_{ij}^{(t)}$ versus conditional node centralities $Z_{it} $; 
(dashed red curve) total degrees $\overline{d}_i   =\sum_t d_i^{(t)} $ versus the sum $\sum_t Z_{it} (\omega)$ of the conditional node centralities; and 
(solid gold curve) degrees $d_i^{(2)} = \sum_j A_{ij}^{(2)}$ in the Ryanair layer versus  $\sum_t Z_{it} (\omega)$. 
}\label{fig:mux_deg_eig}
\end{figure}

In Fig.~\ref{fig:mux_deg_eig}, we illustrate that the eigenvector supracentralities correlate strongly with node degrees. In Fig.~\ref{fig:mux_deg_eig}(a), we show for $\omega=100$ that the airports' conditional centralities, averaged across layers, are correlated strongly with their total (i.e., layer-aggregated) degrees $\overline{d}_i = \sum_{t,j} A_{ij}^{(t)}$ (see the blue `$\times$' marks). We expect this strong correlation for eigenvector supracentrality, as node degree is a first-order approximation of eigenvector centrality in monolayer networks \cite{taylor2011}. 
We also plot the mean conditional centralities versus the number of length-2 paths that emanate from each node (see the red circles).  As expected, this correlation is even stronger, as the number $\sum_{t,j} [A^{(t)}]_{ij}^2$ of length-2 paths is a second-order approximation to eigenvector centrality \cite{taylor2011}.\footnote{As described in \cite{taylor2011}, one can interpret the number of length-$k$ paths as an order-$k$ approximation to the dominant eigenvector of an adjacency matrix when the largest positive (i.e., dominant) eigenvalue has a magnitude that is strictly larger than those of all of the other eigenvalues. Recall that Thm.~\ref{thm:PF} guarantees that this true whenever ${\bf A}$ is nonnegative, irreducible, and aperiodic. (We also note that the aperiodic assumption is invalid for certain classes of networks, such as bipartite networks.) 
Consider the power-method iteration for numerically computing the dominant eigenvector for any adjacency matrix ${\bf A}$ under these assumptions. 
If one initializes the power method with ${\bf 1}$ (a vector of ones), then ${\bf A}^k {\bf 1}$ converges (after normalization) to the dominant eigenvector of  ${\bf A}$.  {Additionally,}
$[{\bf A}^k {\bf 1}]_i$ (i.e., the $i$-th entry of the vector) is equal to the number of length-$k$ paths that emanate from node $i$. \label{footyFOOT}}

In Fig.~\ref{fig:mux_deg_eig}(b), we plot (as a function of $\omega$) the Pearson correlation coefficient $r$ between node degrees and eigenvector centralities for three cases: 
 ({dot-dashed blue} curve) intralayer degrees $d_i^{(t)} = \sum_j A_{ij}^{(t)}$ versus the conditional node centralities $Z_{it} $; 
({dashed red} curve) total degrees $\overline{d}_i   =\sum_t d_i^{(t)} $ versus the sum $\sum_t Z_{it} (\omega)$ of the conditional node centralities; and 
({solid} gold curve) the degrees $d_i^{(2)} = \sum_j A_{ij}^{(2)}$ in the Ryanair layer versus  $\sum_t Z_{it} (\omega)$.  As expected, for 
very small and very large values of $\omega$, the supracentralities correlate strongly with the Ryanair layer and the layer-aggregated network, respectively. Interestingly, for $\omega\approx0.5$, there is a spike in the correlation between the intralayer degrees and conditional node centralities.

In Sec.~SM2 of the Supplementary Materials, we describe the results that we obtain when repeating these computations with PageRank matrices (see Definition~\ref{def:PageRank}). Figure~SM2 gives an interesting contrast to Fig.~\ref{fig:mux_deg_eig}(b). Because PageRank matrices have the same spectral radius, no layer dominates in the limit of small $\omega$, so there is no eigenvector localization (see Thm.~\ref{thm:uncoupled2}). Instead, in Fig.~SM2(b), we observe for small $\omega$ that the conditional centralities correlate strongly with the nodes' intralayer degrees $d_i^{(t)}$.
In Sec.~SM2, we also compare these findings to results for PageRank ``versatility'' \cite{de2015ranking}, a generalization of centrality that attempts to quantify important nodes in a multilayer network that may not be particularly important in any individual layer. Because we use an interlayer-adjacency matrix that encodes all-to-all coupling, the different multiplex generalizations of eigenvector centrality that use the Khatri--Rao matrix product \cite{sola2013eigenvector,romance2015perron,deford2017multiplex,deford2017new} are all equal to each other and are also equivalent to the eigenvector centrality that is associated with the layer-averaged adjacency matrix $\sum_t \mathbf{A}^{(t)}$. Corollary~\ref{corr:aa} states that this centrality is also equivalent to marginal node centrality for eigenvector supracentralities in the limit $\omega\to\infty$.

\subsection{United States Mathematical-Science Program Rankings using a Ph.D. Exchange Network \cite{taylor2017eigenvector}}\label{sec:temp}

We apply our supracentrality framework to study the prestige of U.S. mathematical-science doctoral-granting programs by examining a temporal network that encodes the graduation and hiring of Ph.D. recipients in the mathematical sciences. \new{Specifically, we analyze the temporal network from \cite{taylor2017eigenvector}, which we constructed using} 
data from the Mathematics Genealogy Project \cite{mgp}. As in \cite{taylor2017eigenvector}, we calculate uniformly and diagonally coupled authority scores, such that a university with a high authority score corresponds to an academic authority. A high-authority university produces desirable students, who tend to be hired by other institutions. 

In our study of the Ph.D. exchange network in \cite{taylor2017eigenvector}, we restricted our attention to undirected, adjacent-in-time coupling that is encoded by an interlayer-adjacency matrix $\tilde{\bm{A}}$ with entries $\tilde{A}_{tt'} =1$ if $|t-t'|=1$ and $\tilde{A}_{tt'} =0$ otherwise. In the present study, by contrast, we consider the effects of causality by coupling time layers using a directed chain with  ``layer teleportation''. Specifically, we use an interlayer-adjacency matrix with elements
\begin{equation}\label{eq:tele}
	\tilde{A}_{tt'} = 
\left\{\begin{array}{rl}
	1\,,& t'-t= 1  \\
\gamma \,,&\text{otherwise} \,.\\
\end{array} \right. 
\end{equation}
Teleportation was introduced for PageRank centrality \cite{gleich2014} to allow centrality matrices that are associated with weakly-connected (and even disconnected) networks to satisfy the irreducibility assumptions of Theorems~\ref{thm:PF} and \ref{thm:SC_implies}. Similarly, we use teleportation between layers to satisfy the assumptions of Thm.~\ref{thm:unique}; this ensures that the supracentralities are positive and unique. We differentiate between these two types of teleportation by referring to the original PageRank notion as \emph{node teleportation} and the teleportation in Eq.~\ref{eq:tele} as \emph{layer teleportation}. See \cite{taylor2019supracentrality} for an exploration of
layer teleportation in further detail.\footnote{{Theorems~2 and 3 in \cite{taylor2019supracentrality} restate Thms.~\ref{thm:uncoupled2} and \ref{thm:singular_limit} from an earlier draft of the present paper. We have since improved the clarity of these results by stating that $\lambda_{\textrm{max}}$ and $\tilde{\lambda}_{\textrm{max}}$ are  ``largest positive'' eigenvalues.}}

\begin{figure}[!h]
\centering
\includegraphics[width=\linewidth]{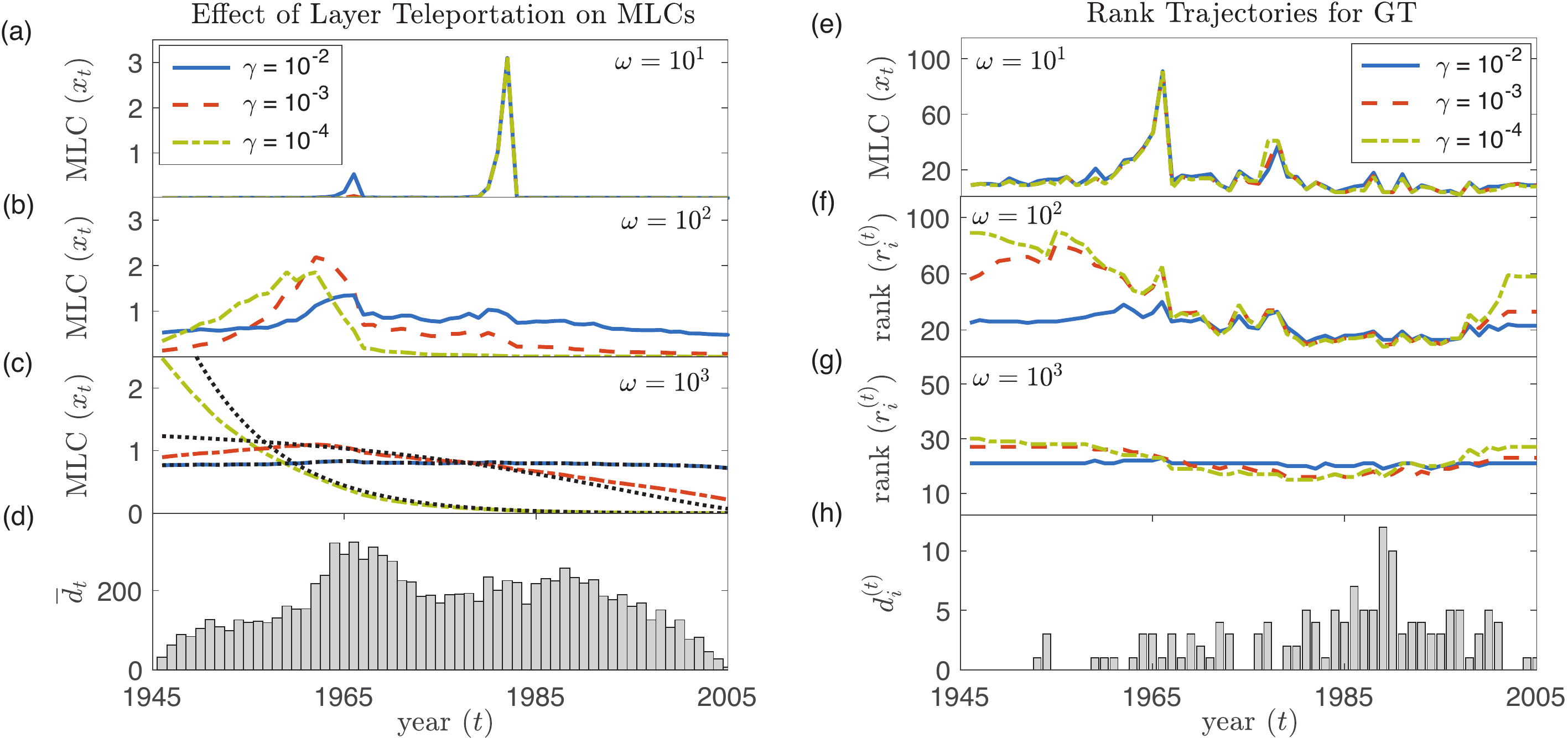}
\vspace{-.6cm}
\caption{
Effect of the layer teleportation parameter $\gamma$ on authority supracentralities in the mathematical-sciences Ph.D. exchange network. 
(a)--(c) MLCs versus the year $t$ for several choices of $\gamma$ and $\omega$. The dotted black curves in panel (c) depict the asymptotic results from
Thm.~\ref{thm:singular_limit}. Specifically, in the $\omega\to\infty$ limit, the MLCs are given by the right dominant eigenvector of $\tilde{\bm{A}}$.
(d) The total number $\overline{d}_t = \sum_{ij}A_{ij}^{(t)}$ of mathematical-sciences Ph.D. recipients in year $t$ who later supervised a graduating Ph.D. student.
(e)--(g) The rank $r_i^{(t)}$ that is associated with the conditional node centrality of Georgia Institute of Technology (GT) for various values of $\gamma$ and $\omega$.
(h) The number $\overline{d}_i^{(t)} = \sum_{j}A_{ij}^{(t)}$ of people who earned a Ph.D. in the mathematical sciences from GT in year $t$ who later supervised a graduating Ph.D. student. 
}
\label{fig:causal_MGP}
\end{figure}

In Fig.~\ref{fig:causal_MGP}, we examine the effect of the \new{layer} teleportation parameter $\gamma$ on \new{authority} supracentralities. In Figs.~\ref{fig:causal_MGP}(a)--(c), we plot the layers' authority MLCs $x_t(\omega)$ (which are given by Definition~\ref{def:marg}) versus the year $t$ for coupling strengths $\omega\in\{10^1,10^{2},10^3\}$ from the weak-coupling, intermediate-coupling, and strong-coupling regimes. See Sec.~\ref{sec:peda} for a description for how we identify coupling regimes. In each panel, we plot the MLCs for three values of the \new{layer} teleportation parameter: $\gamma=10^{-2}$, $\gamma=10^{-3}$, and $\gamma=10^{-4}$. In panel (d), we plot $\overline{d}_t = \sum_{ij}A_{ij}^{(t)}$, which is the total number of people who earned a Ph.D. in the mathematical sciences in year $t$ who later supervised a graduating Ph.D. student. Observe that $t=1966$ is the year with the largest value of $\overline{d}_t$.

In Fig.~\ref{fig:causal_MGP}(a) (i.e., for small $\omega$), we observe eigenvector localization onto time layer $t=1982$, whose associated authority matrix has the largest spectral radius among all layers. For $\gamma=10^{-2}$, we also observe a smaller peak at $t=1966$.
Comparing Fig.~\ref{fig:causal_MGP}(b) to Fig.~\ref{fig:causal_MGP}(a) for $\gamma=10^{-2}$, we observe that the peak near $t=1966$ is more pronounced for $\omega=10^{1}$ than for $\omega=10^{2}$. We observed a similar localization phenomenon in \cite{taylor2017eigenvector} for adjacent-in-time coupling. In Fig.~\ref{fig:causal_MGP}(c), we illustrate behavior that contrasts starkly with our findings in \cite{taylor2017eigenvector}. Specifically, we see that the interlayer-coupling architecture (which changes with the \new{layer} teleportation parameter $\gamma$) has a significant
effect on the strong-coupling limit. By varying $\gamma$, one can tune the extent to which older time layers have larger centrality than newer time layers. When $\gamma=10^{-4}$, for example, one can observe in Fig.~\ref{fig:causal_MGP}(c) that the MLC of time layers appears to decrease rapidly as $t$ increases. Our asymptotic theory in Sec.~\ref{sec:strong} gives an accurate description of this phenomenon. Observe the dotted black curves, which show the right dominant eigenvector vector $\tilde{\bm{v}}^{(1)}$ of $\tilde{\bm{A}}$; we obtain these MLC curves from Thm.~\ref{thm:singular_limit} in the 
$\omega\to\infty$ limit.

In Figs.~\ref{fig:causal_MGP}(e)--(g), we plot the university rank $r_i^{(t)}\in\{1,\dots,N\}$ of Georgia Institute of Technology (GT) that is associated with its conditional centralities for different time layers; we call this its ``rank trajectory''.  Similar to Figs.~\ref{fig:causal_MGP}(a)--(c), these panels show results for $\omega\in\{10,10^2,10^3\}$; in each panel, we again plot results for $\gamma\in\{10^{-2},10^{-3},10^{-4}\}$. In Fig.~\ref{fig:causal_MGP}(h), we plot the number of people who earned a Ph.D. in the mathematical sciences from GT who later supervised a graduating Ph.D. student; observe that this increases starting in the 1960s. Starting in the late 1970s, GT's mathematics program transitioned from being primarily teaching-oriented to being much more research-oriented, with a newly restructured doctoral degree program \cite{DukeGT}.
We used GT in \cite{taylor2017eigenvector} as a case study to illustrate the methods that we developed in that paper. All centrality trajectories for GT that we present in the present paper differ significantly from those in Taylor et al. \cite{taylor2017eigenvector}, who implemented coupled time layers using an undirected chain (i.e., adjacent-in-time coupling with undirected interlayer edges between corresponding universities). In particular, many of the rank trajectories in Figs.~\ref{fig:causal_MGP}(e)--(g) suggest that GT has its highest rank in the 1980s, around the time when GT graduated its largest numbers of Ph.D. students who subsequently supervised their own Ph.D. students.
%
We also see
that the choices for $\omega$ and $\gamma$ influence the centrality trajectory for GT. We observe that $\gamma$ has a larger effect for the intermediate-coupling and strong-coupling regimes (see panels (f) and (g)) of $\omega$ than it does for the weak-coupling regime. Additionally, for the intermediate choices $\gamma=10^{-3}$ and $\omega=10^2$, the rank of GT varies from about 50th to about 15th during when
$t\in\{1945,\ldots, 1985\}$. 

In Sec.~SM3 of the Supplementary Materials, we provide additional results for the Ph.D. exchange data. We show in Table SM.~3 that, although $\gamma$ has a strong effect on the MLCs for large values of $\omega$, it does not seem to have a significant effect on top-ranked schools. 
\new{In Fig.~SM3, we show plots that are similar to  Figs.~\ref{fig:mux_marg_eig} and \ref{fig:mux_deg_eig}, except that the results describe the Ph.D. exchange network, rather than the multiplex airline network.}
%
For the limits of small and large $\omega$, we find that the authority supracentralies correlate with the intralayer degrees of a dominating layer ($t=1966$ in this case) and with the nodes' total degrees, respectively.

\section{Conclusions}\label{sec:discuss}
%
It is important to develop systematic ways of calculating importances in the form of centralities and their generalizations for nodes, edges, and other structures in multilayer networks. In the present paper, we examined centralities that are based on eigenvectors for two popular classes of multilayer networks: (1) multiplex networks, which encode different types of relationships; and (2) temporal networks, in which relationships change in time. We presented a unifying 
linear-algebraic framework that generalizes eigenvector-based centralities, such as PageRank and hub/authority scores, for multiplex and temporal networks.  A key aspect of our approach involves studying joint, marginal, and conditional centralities that one calculates from the dominant eigenvector of a supracentrality matrix, which couples centrality matrices that are associated with individual layers. See \cite{code} for {\sc Matlab} code that computes supracentralities and reproduces the experimental results of our paper.

Our main methodological contribution is the extension of the supracentrality framework of \cite{taylor2017eigenvector}, which previously was restricted to undirected adjacent-in-time coupling, to more general types of interlayer coupling. Our new, more general framework couples layers through an interlayer-adjacency matrix $\tilde{\bm{A}}$, allowing one to study centralities in multilayer networks with a large family of interlayer coupling topologies. We found that the architecture of $\tilde{\bm{A}}$ significantly impacts supracentralities, and we highlighted that some choices are more appropriate than others for different applications. As an example, in Sec.~\ref{sec:temp}, we let $\tilde{\bm{A}}$ encode a directed chain with layer teleportation (see Fig.~\ref{fig:toy1}(b) for a visualization) and studied a temporal network that encodes the graduation and hiring of Ph.D. recipients in the mathematical sciences. Our results on this example contrast sharply with those in \cite{taylor2017eigenvector} because of our different choice of interlayer coupling\footnote{For an in-depth comparison of the effects directed and undirected coupling between time layers, see our recent book chapter \cite{taylor2019supracentrality}. One of our findings in \cite{taylor2019supracentrality} is that coupling layers using a directed chain (which respects the arrow of time) introduces a bias that increases the centralities of node-layer pairs that are associated with the earliest time layers. For example, we illustrated that one can moderate such a bias using a layer teleportation parameter (and we note that it seems fruitful to study multilayer generalizations of so-called ``smart teleportation'' \cite{lambiotte2012}).}.
We also studied a multiplex network that encodes airline transportation in Europe (see Sec.~\ref{sec:mux}), where we found a different mechanism that yields boosts in centrality. Specifically, we observed that nodes that are important in both the large-$\omega$ and small-$\omega$ limits can receive centrality boosts for intermediate values of $\omega$ compared to nodes that are important in just one of the limits.
 For one illustration, see the centrality of the airport LEMB in Table \ref{table:tops} and the black curves in Fig.~\ref{fig:mux_marg_eig}(b).


We explored how different interlayer-coupling architectures (as encoded by $\tilde{\bm{A}}$) and interlayer-coupling strengths (as encoded by $\omega$) influence centralities. Specifically, we identified an interesting interplay between $\tilde{\bm{A}}$ and the architectures of the individual layers in multiplex and temporal networks. To gain insight into this interplay, we performed
singular perturbation theory in the limits of weak and strong interlayer coupling (see Sec.~\ref{sec:limiting}), which lead to layer decoupling and layer aggregation, respectively. We demonstrated that the limiting supracentralies depend on several factors, including the  right and left dominant eigenvectors of $\tilde{\bm{A}}$ and the spectral radii of the layers' centrality matrices, possibly leading to localization of the dominant eigenvector of a supracentrality matrix onto one or more layers (specifically, the ones whose associated centrality matrices have the largest spectral radii). 
\new{We focused on studying the eigenvector that is associated with the dominant eigenvalue of a supracentrality matrix, because we obtain the supracentralities from that vector. It may also be interesting to explore other eigenvectors and eigenvalues using the
perspective of Jordan decompositions \cite{serra2005jordan}.
}


We expect that our results will be useful not only for centrality analysis, but also for 
studying matrices that arise in 
data integration. In the context of our problem, we considered both (i) a set of matrices and (ii) a set of relationships between these matrices. Using a ``supramatrix" framework, we constructed and analyzed a matrix that reflects both (i) and (ii). Our perturbative approach for analyzing dominant eigenspaces in the present paper assumes that the matrices are nonnegative and square, but it is not limited to matrices that encode network data. Consequently, we expect our findings to also be insightful in other scenarios that involve combining matrices of data into larger matrices.


\new{One can also use a supracentrality framework to study dynamical processes on multiplex networks. For example, multiplex Markov chains were defined recently \cite{taylor2020multiplex} using a formulation that was inspired by supracentrality. Such chains can outperform other diffusion models when assessing the importances of nodes in a multiplex network.}

\appendix

\section{Proof of Theorem~\ref{thm:unique}}\label{app:unique}
%

To prove the uniqueness and positivity of $\mathbbm{v}(\omega)$, we use the Perron--Frobenius theorem for nonnegative matrices (see Thm.~\ref{thm:PF}) \cite{bapat1997}. To satisfy the assumptions of Thm.~\ref{thm:PF}, we must show that the matrix $\mathbb{C}(\omega)$ is nonnegative and irreducible under our two assumptions: (i) $\tilde{\bm{A}}$ is nonnegative and irreducible; and (ii) the sum $\sum_t \mathbf{C}^{(t)}$ is an irreducible nonnegative matrix.  By construction, the entries in $\mathbb{C}(\omega)$ are nonnegative, so we only need to prove irreducibility. Because the matrix $\mathbb{C}(\omega)$ is nonnegative, we can interpret it as a weighted adjacency matrix that encodes (possibly) directed and weighted edges between node-layer pairs $\{(i,t)\}$ for $i\in\mathcal{V}=\{1,\dots,N\}$ and $t\in\{1,\dots,T\}$. We will show that the network that is associated with the adjacency matrix $\mathbb{C}(\omega)$ is strongly connected, which implies that it is irreducible.

We start with two observations. First, $\tilde{\bm{A}}$ describes an adjacency matrix for a strongly connected network. Let $\tilde{L}<\infty$ denote the diamater of this network. For any node $k$ and any two layers, $t$ and $t'$, it follows that there exists a path from node-layer pair $(k,t)$ to node-layer pair $(k,t')$ in the network that is associated with the adjacency matrix $\mathbb{C}(\omega)$. The length of this path is at most $\tilde{L}$. Second, because the matrix $\overline{ \mathbf{C}^{(t)}} = T^{-1}\sum_t \mathbf{C}^{(t)}$ is irreducible and nonnegative, we can interpret it as an adjacency matrix for a strongly connected network. Let ${L}<\infty$ denote the diameter of this network. For any two nodes, $i$ and $j$, of this network, it follows that there exists a path of length $l\le L$ from $i$ to $j$. We denote the path by a sequence $\mathcal{P}(i,j)=\{k_0,k_1,\dots, k_{l-1},k_l\}$ of nodes from $i=k_0$ to $j=k_l$. We also identify a sequence $\{t_1,t_2,\dots,t_l\}$ of layers, such that the entry $[\mathbf{C}^{(t_j)}]_{k_{j-1},k_j}$ in matrix $\mathbf{C}^{(t_j)}$ is positive. For any $j$, there must exist at least one matrix $\mathbf{C}^{(t_j)}$ for which the  $(k_{j-1},k_j)$-th entry is positive, because the $(k_{j-1},k_j)$-th entry in $\overline{ \mathbf{C}^{(t)}}$ is positive (i.e., because $(k_{j-1},k_j)$ is an edge) and $\overline{ \mathbf{C}^{(t)}}$ is a sum of nonnegative matrices. 

For any two node-layer pairs, $(i,s)$ and $(j,t)$, we now prove that there exists a path, with a length of at most $\tilde{L}L$, from $(i,s)$ to $(j,t)$ in the network that is associated with the matrix $\mathbb{C}(\omega)$. We do this by explicitly constructing such a path. We first identify a path $\mathcal{P}(i,j)$ from $i$ to $j$ in the network that is associated with $\overline{ \mathbf{C}^{(t)}}$. Consider the following sequence of node-layer pairs: 
\begin{equation}
	\{(k_0,t_0),(k_1,t_1),(k_2,t_2),\dots,(k_{l-1},t_{l-1}),(k_l,t_l)\}\,,
\label{eq:booboo}
\end{equation}
where $l\le {L}$; we define $k_j$ and $t_j$ as above; and $k_0=i$, $t_0=s$, $k_l=j$, and $t_l=t$. By definition, the $(k_{j-1},k_j)$-th entry in $\mathbf{C}^{(t_j)}$ is positive for each $j$, implying that the network that is associated with the matrix $\mathbb{C}(\omega)$ has an edge from $(k_{j-1},t_j)$ to $(k_{j},t_j)$. We construct a path from $(i,s)$ to $(j,t)$ by taking the sequence in 
\eqref{eq:booboo} and inserting a path from each term in the sequence to the next term. That is, we insert a path from $(k_0,t_0)$ to $(k_0,t_1)$ using only node-layer pairs that involve node $k_0$. The length of this path is at most $\tilde{L}$. Additionally, from our definition of the path $P(i,j)$, we see that there exists an edge from $(k_0,t_1)$ to $(k_1,t_1)$. We then insert a path, whose length is also at most $\tilde{L}$, from $(k_1,t_1)$ to $(k_1,t_2)$ using only node-layer pairs that involve node $k_1$. There also exists an edge from $(k_1,t_2)$ to $(k_2,t_2)$, and so on. We repeat this process until finally we insert a path from  $(k_{l-1},t_{l-1})$ to $(k_{l-1},t_{l})$ using only node-layer pairs that involve node $k_{l-1}$, and we note that there exists an edge from $(k_{l-1},t_{l})$ to $(k_l,t_l)$. 
Each of these paths exists because the network that is associated with $\tilde{\bm{A}}$ is strongly connected, and each of these paths has a length of at most $\tilde{L}$.

This construction yields a path from any node-layer pair to any other node-layer pair in the network that is associated with $\mathbb{C}(\omega)$, which proves that the network is strongly connected. Using our construction, we have also obtained an upper bound for the network's diameter of
$\tilde{L}L$. Because $\mathbb{C}(\omega)$ corresponds to a strongly connected network, it is irreducible and nonnegative by the Perron--Frobenius theorem for nonnegative matrices, so the right dominant eigenvector $\mathbbm{v}(\omega)$ is unique and positive. Consequently, the entries $\{W_{it}(\omega)\}$, $\{x_i(\omega)\}$, and $\{\hat{x}_t(\omega)\}$ are also unique and positive. Because these entries are positive, it follows in turn that $ \{Z_{it}(\omega)\}$ and $ \{\hat{Z}_{it}(\omega)\}$ are positive and finite.

Finally, if $\mathbb{C}(\omega)$ is also aperiodic, then Thm.~\ref{thm:PF} states that the largest positive eigenvalue of $\mathbb{C}(\omega)$ is larger in magnitude than the other eigenvalues.

\section{Proof of \new{Lemma}~\ref{thm:uncoupled}}\label{app:uncoupled}
\quad We show that each eigenvalue--eigenvector pair of ${\bf C}^{(t)}$ yields an eigenvalue--eigenvector pair of $\mathbb{C}(0)$.
{Consider the matrix--vector multiplication $\new{\mathbb{C}(0)} \mathbbm{v}^{(i,t)}$, and let $k=\new{p}~\text{mod}(N)$ and $t'= \lceil \new{p}/jN \rceil$. 
We write
\begin{align}
	[\new{\mathbb{C}(0)} \mathbbm{v}^{(i,t)}]_\new{p} = \sum_{(k,t')} C_{jk}^{(t)}v_k^{(i,t')}\delta_{tt'} = \mu_i^{(t)} v_j^{(i,t')}\delta_{tt'}  =\mu_i^{(t)} [\mathbbm{v}^{(i,t)}]_\new{p}  \,,
\end{align}
where we use the notation $[\hat{\mathbb{C}} \mathbbm{v}^{(i,t)}]_\new{p}$ to denote the $\new{p}$-th entry of the vector $\new{\mathbb{C}(0)} \mathbbm{v}^{(i,t)}$. 
This implies that $\new{\mathbb{C}(0)} \mathbbm{v}^{(i,t)} =\mu_i^{(t)} \mathbbm{v}^{(i,t)}$, so $\mu_i^{(t)}$ is an eigenvalue of $\new{\mathbb{C}(0)}$ with right eigenvector $\mathbbm{v}^{(i,t)}$. Similarly, $\new{\mathbb{C}(0)}^\new{*} \mathbbm{u}^{(i,t)} =\mu_i^{(t)} \mathbbm{u}^{(i,t)}$, so $\mathbbm{u}^{(i,t)}$ is the associated left eigenvector. 
} 

\section{Proof of Theorem~\ref{thm:uncoupled2}}\label{app:uncoupled2}
%
Equation~\eqref{eq:lim_uv} follows from \new{Lemma}~\ref{thm:uncoupled} and \new{Remark~\ref{remark:ahh}}. Let $\new{\mathcal{T}}$ be \new{the} set of layer indices $\{t\}$ for which $\mu_1^{(t)}= \lambda_{\text{max}}(0)$, and define \new{$\mathcal{S} = \text{span}\left(\{ \mathbbm{v}^{(1,t)} \}_{t\in\new{\mathcal{T}}}\right)$ as the span of the associated eigenvectors (i.e., as the right dominant eigenspace). Because of the continuity of ``eigenprojections'' (i.e., projections onto eigenspaces) \cite{kato2013perturbation} (see Ch.~2), the dominant eigenvector $\mathbbm{v}^{(1)}(\omega)$ converges to the dominant eigenspace $\mathcal{S}$ of $\mathbb{C}(0)$. Similarly, when $\omega \to 0^+$, the  left dominant eigenvector of $\mathbb{C}(\omega)$ converges to the  left dominant eigenspace of $\mathbb{C}(0)$. Given these observations, we only need to
}
prove that the constants $\{\alpha_\new{t}\}$ and $\{\beta_\new{t}\}$ satisfy Eqs.~\eqref{eq:baah}.





We expand $\lambda_\textrm{\rm{max}}(\omega)$, $\mathbbm{u}^{(1)}(\omega)$, and $\mathbbm{v}^{(1)}(\omega)$ for small $\omega$ to obtain order-$k$ approximations:
\begin{align}\label{these}
	\lambda_\textrm{\rm{max}}(\omega) &=  \sum_{j=0}^k\omega^j\lambda_j + \mathcal{O}(\omega^{k+1})\,, \nonumber\\
	\mathbbm{v}^{(1)}(\omega) &=  \sum_{j=0}^k\omega^j\mathbbm{v}_j + \mathcal{O}(\omega^{k+1}) \,, \nonumber\\
		\mathbbm{u}^{(1)}(\omega) &=  \sum_{j=0}^k\omega^j\mathbbm{u}_j + \mathcal{O}(\omega^{k+1})\, .
\end{align}
We use superscripts to indicate powers of $\omega$ in the terms in the expansion, and we use subscripts for the terms that are multiplied by a power of $\omega$. Note that $\lambda_0$, $\mathbbm{v}_0$, and $\mathbbm{u}_0$,  respectively, denote the dominant eigenvalue and its corresponding right and left eigenvectors in the $\omega\to0^+$ limit.  Successive terms in the expansions \eqref{these} \new{involve} higher-order derivatives, and we assume that each term has appropriate smoothness of these functions. 


Our strategy is to develop consistent solutions to
$ {\mathbb{C}}(\omega)^\new{*} {\mathbbm{u}}^{(1)}(\omega) = \lambda_{\rm{max}}(\omega) {\mathbbm{u}}^{(1)}(\omega)$ and
$ {\mathbb{C}}(\omega) {\mathbbm{v}}^{(1)}(\omega) = \lambda_{\rm{max}}(\omega) {\mathbbm{v}}^{(1)}(\omega)$
 for progressively larger values of $k$. Let's consider the equation for the right eigenvector. Starting with the first-order approximation, we insert $\lambda_{\rm{max}}(\omega) \approx \lambda_{0} + \omega \lambda_{1}$ and $\mathbbm{v}^{(1)}(\omega) \approx \mathbbm{v}_0 + \omega \mathbbm{v}_1 $ into Eq.~\eqref{eq:eig_eq} and collect the zeroth-order and first-order terms in $\omega$ to obtain
\begin{align}
	\left(\lambda_0\mathbb{I} -  \hat{\mathbb{C}}\right)\mathbbm{v}_0 
		&= 0 \,, \label{eq:firstaagag1} \\
		\left(\lambda_0\mathbb{I} - \hat{\mathbb{C}}\right)\mathbbm{v}_1 
		&= \left(\hat{\mathbb{A}} - \lambda_1\mathbb{I}\right)	\mathbbm{v}_0 \label{eq:adga} \,,
\end{align}
where $\mathbb{I}$ is the $NT\times NT$ identity matrix. Equation~\eqref{eq:firstaagag1} corresponds to the system that is described by \new{Lemma}~\ref{thm:uncoupled}, implying that the operator $\lambda_0\mathbb{I} - \hat{\mathbb{C}}$ is singular and has a $\new{|\new{\mathcal{T}}|}$-dimensional null space, where \new{$\new{\mathcal{T}} = \{t:\mu_t = \max_t \mu_t \}$ is the set} of centrality matrices ${\bm C}^{(t)}$ whose largest eigenvalue is equal to the maximum eigenvalue. In particular, $\max_t \mu_t = \lambda_0 = \lambda_{\rm{max}}(0)$, and the dominant eigenvectors have  \new{a general solution of the form}
form
\begin{equation}
  	 \mathbbm{v}_0 =\sum_{t} \alpha_{t}  \mathbbm{v}^{(1,t)}\,, ~~ 
	 \mathbbm{u}_0 =\sum_{t} \beta_t  \mathbbm{u}^{(1,t)} \,,  \label{eq:vasgas}
\end{equation}
where $\alpha_{t}$ and $\beta_{t}$ are constants that satisfy $1= \sum_t \alpha_{t}^2=\sum_t \beta_{t}^2$ to ensure that $\| \bm{\alpha}\|=\| \bm{\beta}\|=1$. (See \new{Lemma}~\ref{thm:uncoupled} for definitions of $ \mathbbm{u}^{(1,t)}$ and $ \mathbbm{v}^{(1,t)}$.)



To determine the vectors $\bm{\alpha} = [\alpha_1,\dots,\alpha_T]^\new{*}$ and $\bm{\beta} = [\beta_1,\dots,\beta_T]^\new{*}$ of  constants that uniquely determine $\mathbbm{u}_0$ and $\mathbbm{v}_0$, we use Eq.~\eqref{eq:adga} to seek a solvability condition for the first-order terms. We use the fact that the left null space of $\lambda_0\mathbb{I} - \hat{ \mathbb{C}}$ is the span of $\{ \mathbbm{u}^{(1,t)}\}$ to see that $[\mathbbm{u}^{(1,t)}]^\new{*} \left(\lambda_0\mathbb{I} - \hat{\mathbb{C}}\right)\mathbbm{v}_1 = 0$ for any  $t$. We  left-multiply  Eq.~\eqref{eq:adga} by $[\mathbbm{u}^{(1,t)}]^\new{*} $ and simplify to obtain
\begin{align}
	[\mathbbm{u}^{(1,t)}]^\new{*} \hat{\mathbb{A}} \mathbbm{v}_0  &=  \lambda_1 [\mathbbm{u}^{(1,t)}]^\new{*}  \mathbbm{v}_0 \,.\label{eq:bagaga}
\end{align}
Using the solution of $\mathbbm{v}_0$ in Eq.~\eqref{eq:vasgas}, we obtain 
\begin{align}
	\sum_{t'} \alpha_{t'} [\mathbbm{u}^{(1,t)}]^\new{*} \hat{\mathbb{A}}  \mathbbm{v}^{(1,t')}
		&=  \lambda_1 \sum_{t'}  \alpha_{t'}  [\mathbbm{u}^{(1,t)}]^\new{*}  \mathbbm{v}^{(1,t')} \nonumber\\
		&=  \lambda_1  \langle {\bf u}^{(1,t)} , {\bf v}^{(1,t)}\rangle  \alpha_t\, ,\label{eq:eig_aadgag}
\end{align}
which uses $  [\mathbbm{u}^{(1,t)}]^\new{*}  \mathbbm{v}^{(1,t)} = \langle \bm{u}^{(1,t)} , \bm{v}^{(1,t')}\rangle\delta_{tt'} $, where $\delta_{ij}$ is the Kronecker delta. We simplify the left-hand side of Eq.~\eqref{eq:eig_aadgag} to obtain
\begin{align}
	\sum_{t'} \alpha_{t'} [\mathbbm{u}^{(1,t)}]^\new{*} \hat{\mathbb{A}}  \mathbbm{v}^{(1,t')}
		&=  \sum_{t'} \alpha_{t'}  {\tilde{A}_{tt'}} [\mathbbm{u}^{(1,t)}]^\new{*} [ {{\bf e}}^{(t')}\otimes{\bf v}^{(1,t')} ] \nonumber\\
		&=  \sum_{t'} \alpha_{t'}  {\tilde{A}_{tt'}} [{{\bf e}}^{(t)}\otimes{\bf u}^{(1,t)}]^\new{*} [ {{\bf e}}^{(t')}\otimes{\bf v}^{(1,t')} ] \nonumber\\
		&=  \sum_{t'} \alpha_{t'}  {\tilde{A}_{tt'}} \langle {\bf u}^{(1,t)} ,  {\bf v}^{(1,t')}\rangle  \,. \label{eq:eig_aadgag2}
\end{align}
The last expression follows from the relations $\hat{\mathbb{A}}=\tilde{\bm{A}}\otimes \bf{I}$, $\mathbbm{v}^{(1,t')} =  {{\bf e}}^{(t')}\otimes{\bf v}^{(1,t')}$, and $\mathbbm{u}^{(1,t)} =  {{\bf e}}^{(t)}\otimes{\bf u}^{(1,t)}$, where we recall that ${{\bf e}}^{(t)}$ is a unit vector that consists of zeros
in all entries except for entry $t$ (which is a $1$). We equate the expressions~\eqref{eq:eig_aadgag2} and \eqref{eq:eig_aadgag}
and divide by $\langle {\bf u}^{(1,t)} , {\bf v}^{(1,t)}\rangle$ to obtain the equation
\begin{align}
	\sum_{t'}   {\tilde{A}_{tt'}} \frac{\langle {\bf u}^{(1,t)}, {\bf v}^{(1,t')} \rangle}{\langle {\bf u}^{(1,t)} , {\bf v}^{(1,t)}\rangle}  \alpha_{t'} &=  \lambda_1   \alpha_t\, 
\end{align}
for the right dominant eigenvalue. One proceeds analogously to obtain a \new{similar} equation for the   left dominant eigenvector,
and together these two eigenvector equations yield Eq.~\eqref{eq:baah}.


\section{Proof of \new{Lemma}~\ref{thm:singularity}}\label{app:singularity}
%
Examining $\tilde{\mathbb{C}}(\epsilon) $, which is given by Eq.~\eqref{eq:newC}, yields (using any matrix norm)
$\|\tilde{\mathbb{C}}(\epsilon)  -  \hat{\mathbb{A}}\| = \epsilon \|\hat{\mathbb{C}}\| \to0^+$ 
as $\epsilon\to 0^+$, implying that
$\tilde{\mathbb{C}}(0^+)  = \hat{\mathbb{A}} =  \tilde{\bm{A}} \otimes \mathbf{I}$. 
Using the stride permutation $\mathbb{P}$ that is defined by Eq.~\eqref{eq:stride}, we 
write
\begin{equation}\label{eq:perm2}
 	\mathbb{P}^\new{*} ( \tilde{\bm{A}} \otimes \mathbf{I} )\mathbb{P} = \mathbf{I} \otimes \tilde{\bm{A}}  
		=\left[ \begin{array}{ccc} 
			 \tilde{\bm{A}} & 0 & \cdots\\ 
			0& \tilde{\bm{A}}  &  \ddots\\ 
			 \vdots &   \ddots&\ddots\\
			 \end{array}
		\right] \, .
\end{equation}
Because $\mathbf{I} \otimes \tilde{\bm{A}}$ is block diagonal and each diagonal block is identical, it follows that the spectrum of $\mathbf{I} \otimes \tilde{\bm{A}}$ is identical to that of $\tilde{\bm{A}}$ (although the eigenvalues need to repeat an appropriate number of times), and one can obtain the eigenvectors of the former as functions of the eigenvectors of $\tilde{\bm{A}}$.

Let $\{\tilde{\mu}_t\}$ denote the eigenvalues of $\tilde{\bm{A}}$, and let $\tilde{\bm{v}}^{(t)}$ and $\tilde{\bm{u}}^{(t)}$ denote their corresponding   right and left  eigenvectors, respectively. We now illustrate that $\tilde{\mathbbm{v}}^{(t,j)} = \tilde{{\bf e}}^{(j)} \otimes  {\bf v}^{(t)}$ and $\tilde{\mathbbm{u}}^{(t,j)} = \tilde{{\bf e}}^{(j)} \otimes  {\bf u}^{(t)}$ are right and left  eigenvectors of $ \mathbf{I} \otimes \tilde{\bm{A}}$. With $\new{p}\in\{1,\dots,NT\}$, we define $t=\new{p}~\text{mod}(T)$ and $k=  \lceil t/\new{p}T \rceil$ and obtain
\begin{align}\label{eq:u12}
	[(\mathbf{I} \otimes \tilde{\bm{A}}) \hat{\mathbbm{v}}^{(t,j)}]_\new{p} 
	&= \sum_{t',k'} \tilde{A}_{tt'}\delta_{kk'}  { \tilde{v}}_{t'}^{(t)} \delta_{k'j} \nonumber\\
	&= \sum_{t',k'} \tilde{A}_{tt'}  { \tilde{v}}_{t'}^{(t)} \delta_{kj}  \nonumber\\
	&= \sum_{t'} \tilde{A}_{tt'}  { \tilde{v}}_{t'}^{(t)} \delta_{kj}\nonumber\\
	&= \tilde{\mu}_{t} { \tilde{v}}_{t'}^{(t)} \delta_{kj} \nonumber\\
	&= \tilde{\mu}_{t}  [\hat{\mathbbm{v}}^{(t,j)} ]_\new{p}\,.
\end{align}
One can show similarly that $(\mathbf{I} \otimes \tilde{\bm{A}})^\new{*} \tilde{\mathbbm{u}}^{(t,j)} = \tilde{\mu}_{t}  \tilde{\mathbbm{u}}^{(t,j)} $, illustrating that $\tilde{\mathbbm{v}}^{(t,j)}$ and $\tilde{\mathbbm{u}}^{(t,j)}$ are right and left eigenvectors that are associated with the eigenvalue $\tilde{\mu}_t$ of $\mathbf{I} \otimes \tilde{\bm{A}}$. This implies that $\mathbb{P}^\new{*} ( \bm{A} \otimes \mathbf{I} )\mathbb{P}  \tilde{\mathbbm{v}}^{(t,j)} = \mu_{t}  \tilde{\mathbbm{v}}^{(t,j)} $, and left-multiplication by $\mathbb{P} $ gives
\begin{align}\label{eq:evec_fin}
	( \tilde{\bm{A}} \otimes \mathbf{I} ) [\mathbb{P}  \tilde{\mathbbm{v}}^{(t,j)}] = \tilde{\mu}_t [\mathbb{P}  \tilde{\mathbbm{v}}^{(t,j)}]\, .
\end{align}

By repeating this procedure using $( \tilde{\bm{A}} \otimes \mathbf{I} )^\new{*}$ (instead of $  \tilde{\bm{A}} \otimes \mathbf{I}  $), one can also show that $( \tilde{\bm{A}} \otimes \mathbf{I} )^\new{*} [\mathbb{P}  \tilde{\mathbbm{u}}^{(t,j)}] = \tilde{\mu}_t [\mathbb{P}  \tilde{\mathbbm{u}}^{(t,j)}]$. Taken together, this expression and Eq.~\eqref{eq:u12}
imply that  $\mathbb{P}  \tilde{\mathbbm{v}}^{(t,j)}$ and $\mathbb{P}  \tilde{\mathbbm{u}}^{(t,j)}$   are right and left eigenvectors of $ \hat{\mathbb{A}} =  \tilde{\bm{A}} \otimes \mathbf{I}$ that are associated with the eigenvalue $\tilde{\mu}_t$. However, for a given value of $t$ (and assuming that the eigenvalues $\{\tilde{\mu}_t\}$ are simple), there are    {$N$ orthogonal}  right eigenvectors $\{\mathbb{P}  \tilde{\mathbbm{v}}^{(t,j)} \}$ and $N$ orthogonal left eigenvectors $\{\mathbb{P}  \tilde{\mathbbm{u}}^{(t,j)} \}$ for $j\in\{1,\dots,N\}$. It follows that each eigenvalue $\tilde{\mu}_t$ of $\tilde{\mathbb{A}}$ has multiplicity $N$ and associated $N$-dimensional right and left eigenspaces.

\section{Proof of Theorem~\ref{thm:singular_limit}}\label{app:singular_limit}
%
We expand $\tilde{\lambda}_{\rm{max}}(\epsilon)$,  $\tilde{\mathbbm{v}}^{(1)}(\epsilon)$, and  $\tilde{\mathbbm{u}}^{(1)}(\epsilon)$  for small $\epsilon$ to obtain order-$k$ approximations:
\begin{align}\label{these2}
	\tilde{\lambda}_{\rm{max}}(\epsilon) &=  \sum_{j=0}^k\epsilon^j \tilde{\lambda}_j + \mathcal{O}(\epsilon^{k+1})\,, \nonumber\\
	\tilde{\mathbbm{v}}^{(1)}(\epsilon) &=  \sum_{j=0}^k\epsilon^j\tilde{\mathbbm{v}}_j +  \mathcal{O}(\epsilon^{k+1}) \,, \nonumber\\
	\tilde{\mathbbm{u}}^{(1)}(\epsilon) &=  \sum_{j=0}^k\epsilon^j\tilde{\mathbbm{u}}_j +  \mathcal{O}(\epsilon^{k+1}) \, .
\end{align}
We use superscripts to indicate powers of $\epsilon$ in the terms in the expansion and subscripts for the terms that are multiplied by $\epsilon^j$. Note that $\tilde{\lambda}_0$, $\tilde{\mathbbm{v}}_0$, and $\tilde{\mathbbm{u}}_0$, respectively, indicate the dominant eigenvalue and its corresponding right and left \new{dominant} eigenvectors in the $\epsilon\to0^+$ limit.  Successive terms in the expansions \eqref{these2} \new{involve} higher-order derivatives, and we assume that each term has appropriate smoothness of these functions. 


Our strategy is to develop consistent solutions to both of the eigenvalue equations, $\tilde{\mathbb{C}}(\epsilon)^\new{*}  \tilde{\mathbbm{u}}(\epsilon) = \tilde{\lambda}_{\rm{max}}(\epsilon) \tilde{\mathbbm{u}}(\epsilon)$ and $\tilde{\mathbb{C}}(\epsilon) \tilde{\mathbbm{v}}(\epsilon) = \tilde{\lambda}_{\rm{max}}(\epsilon) \tilde{\mathbbm{v}}(\epsilon)$, for progressively larger values of $k$. Let's consider the equation for the right eigenvector. Starting with the first-order approximation, we insert $\tilde{\lambda}_{\rm{max}}(\epsilon) \approx \tilde{\lambda}_{0} + \epsilon \tilde{\lambda}_{1}$ and $\tilde{\mathbbm{v}}(\epsilon) \approx \tilde{\mathbbm{v}}_0 + \epsilon \tilde{\mathbbm{v}}_1 $ into Eq.~\eqref{eq:eig_eq} and collect the zeroth-order and first-order terms in $\epsilon$ to obtain
\begin{align}
	\left(\tilde{\lambda}_0\mathbb{I} -  \hat{\mathbb{A}}\right) \tilde{\mathbbm{v}}_0 
		&= 0 \,, \label{eq:first_1} \\
		\left(\tilde{\lambda}_0\mathbb{I} - \hat{\mathbb{A}}\right) \tilde{\mathbbm{v}}_1 
		&= \left(\hat{\mathbb{C}} - \tilde{\lambda}_1\mathbb{I}\right) \tilde{\mathbbm{v}}_0 \label{eq:first_2} \,,
\end{align}
where $\mathbb{I}$ is the $NT\times NT$ identity matrix. Equation~\eqref{eq:first_1} corresponds to the system that is described by Lemma~\ref{thm:uncoupled}, implying that the operator $\tilde{\lambda}_0\mathbb{I} - \hat{\mathbb{A}}$ is singular and has an $N$-dimensional null space. 
(This is the dominant eigenspace of $\hat{\mathbb{A}}$.) Specifically, Eq.~\eqref{eq:first_1} has a general solution of the form
\begin{align}	 \label{eq:v0}
	 \tilde{\mathbbm{v}}_0 &=\sum_j \tilde{\alpha}_j \mathbb{P} \tilde{\mathbbm{v}}^{(1,j)} \,,  \nonumber\\
	\tilde{\lambda}_0 &= \max_t \tilde{\mu}_t\,,  
\end{align}
where $\tilde{\alpha}_{i}$ are constants that satisfy $\sum_i \tilde{\alpha}_{i}^2=1$ (which implies that $\|\tilde{\mathbbm{v}}_0\|_2 =1$). Additionally,
\begin{align}
	 \tilde{\mathbbm{u}}_0 &=\sum_j \tilde{\beta}_j \mathbb{P} \tilde{\mathbbm{u}}^{(1,j)} \,,  \label{eq:v0_b}
\end{align}
where $\sum_i \tilde{\beta}_{i}^2=1$ (which implies that $\|\tilde{\mathbbm{u}}_0\|_2 =1$).

To determine the vectors $\tilde{\bm{\alpha}} = [\tilde{\alpha}_1,\dots,\tilde{\alpha}_N]^\new{*}$ and $\tilde{\bm{\beta} }= [\tilde{\beta}_1,\dots,\tilde{\beta}_N]^\new{*}$ of  constants that uniquely determine $\tilde{\mathbbm{u}}_0$ and $\tilde{\mathbbm{v}}_0$, we seek a solvability condition for the first-order terms. Using the fact that the left null space of $\tilde{\lambda}_0\mathbb{I} - \hat{ \mathbb{A}}$ is the span of $\{ \mathbb{P}\tilde{\mathbbm{u}}^{(1,i)} \}$, it follows that $[\tilde{\mathbbm{u}}^{(1,i)}]^\new{*}\mathbb{P}^\new{*}\left(\tilde{\lambda_0}\mathbb{I} - \hat{\mathbb{A}}\right)\tilde{\mathbbm{v}}_1 = 0$. Therefore, we left-multiply Eq.~\eqref{eq:first_2} by $[\tilde{\mathbbm{u}}^{(1,i)}]^\new{*}\mathbb{P}^\new{*}$ and simplify to obtain 
\begin{align}
	[\tilde{\mathbbm{u}}^{(1,i)}]^\new{*}\mathbb{P}^\new{*}\hat{\mathbb{C}} \tilde{\mathbbm{v}}_0  &=  \tilde{\lambda}_1 [\tilde{\mathbbm{u}}^{(1,i)}]^\new{*}\mathbb{P}^\new{*} \tilde{\mathbbm{v}}_0 \,.\label{eq:base0}
\end{align}

Using the solution of $\tilde{\mathbbm{v}}_0$ in Eq.~\eqref{eq:v0}, we obtain 
\begin{align}
	\sum_j \tilde{\alpha}_j [\mathbbm{u}^{(1,i)}]^\new{*}\mathbb{P}^\new{*} \hat{\mathbb{C}}  \mathbb{P} \tilde{\mathbbm{v}}^{(1,j)}
		&=  \tilde{\lambda}_1 \sum_j \tilde{\alpha}_j  [\tilde{\mathbbm{u}}^{(1,i)}]^\new{*}\mathbb{P}^\new{*} \mathbb{P}\tilde{\mathbbm{v}}^{(1,j)}  \nonumber\\
		&=  \tilde{\lambda}_1 \sum_j \tilde{\alpha}_j  [\tilde{\mathbbm{u}}^{(1,i)}]^\new{*} \tilde{\mathbbm{v}}^{(1,j)}  \nonumber\\
		&= \tilde{\lambda}_1  \langle \tilde{\bm{u}}^{(1)} , \tilde{\bm{v}}^{(1)}\rangle    \tilde{\alpha}_i \,,\label{eq:eig_a0}
\end{align}
because $\mathbb{P}^\new{*}\mathbb{P}=\mathbb{P}\mathbb{P}^\new{*}=\mathbb{I}$ and $[\tilde{\mathbbm{u}}^{(1,i)}]^\new{*} \tilde{\mathbbm{v}}^{(1,j)} = \langle \tilde{\bm{u}}^{(1)} , \tilde{\bm{v}}^{(1)}\rangle\delta_{ij}$, where $\delta_{ij}$ is the Kronecker delta. We divide 
Eq.~\eqref{eq:eig_a0} by $ \langle \tilde{\bm{u}}^{(1)} , \tilde{\bm{v}}^{(1)}\rangle$ to obtain an $N$-dimensional eigenvalue equation for the dominant eigenvector $\tilde{\bm{\alpha}}$. One can implement the analogous steps for the equations for left dominant  eigenvector, and together these two eigenvector equations yield Eq.~\eqref{eq:alpha_beta}.

\bibliographystyle{siam}
\bibliography{centrality_bib-new6}

\end{document}


\maketitle


%
In this supplement, we provide more information about our study of our pedagogical synthetic network (see Sec.~\ref{sec:ped}), the multiplex airline network (see Sec.~\ref{sec:euro}), and the Ph.D. exchange network (see Sec.~\ref{sec:phd}).

\section{Extended Study of Our Pedagogical Example}\label{sec:ped}
%
In this section, we present an extended study of our numerical experiments (see Sec.~3.3 of the main text) using 
the multiplex network in Fig.~1(a) of the main text. Recall that the interlayer-coupling strength between layers $3$ and $4$ in this network differs from that of the other interlayer couplings. (See the dashed lines in Fig.~1(a).) In Sec.~3.3, we set $\tilde{A}_{34}=\tilde{A}_{43}=0.01$ and $\tilde{A}_{tt'}=1$ for the other interlayer couplings.

\begin{figure}[h!]
\centering
\includegraphics[width=.45\linewidth]{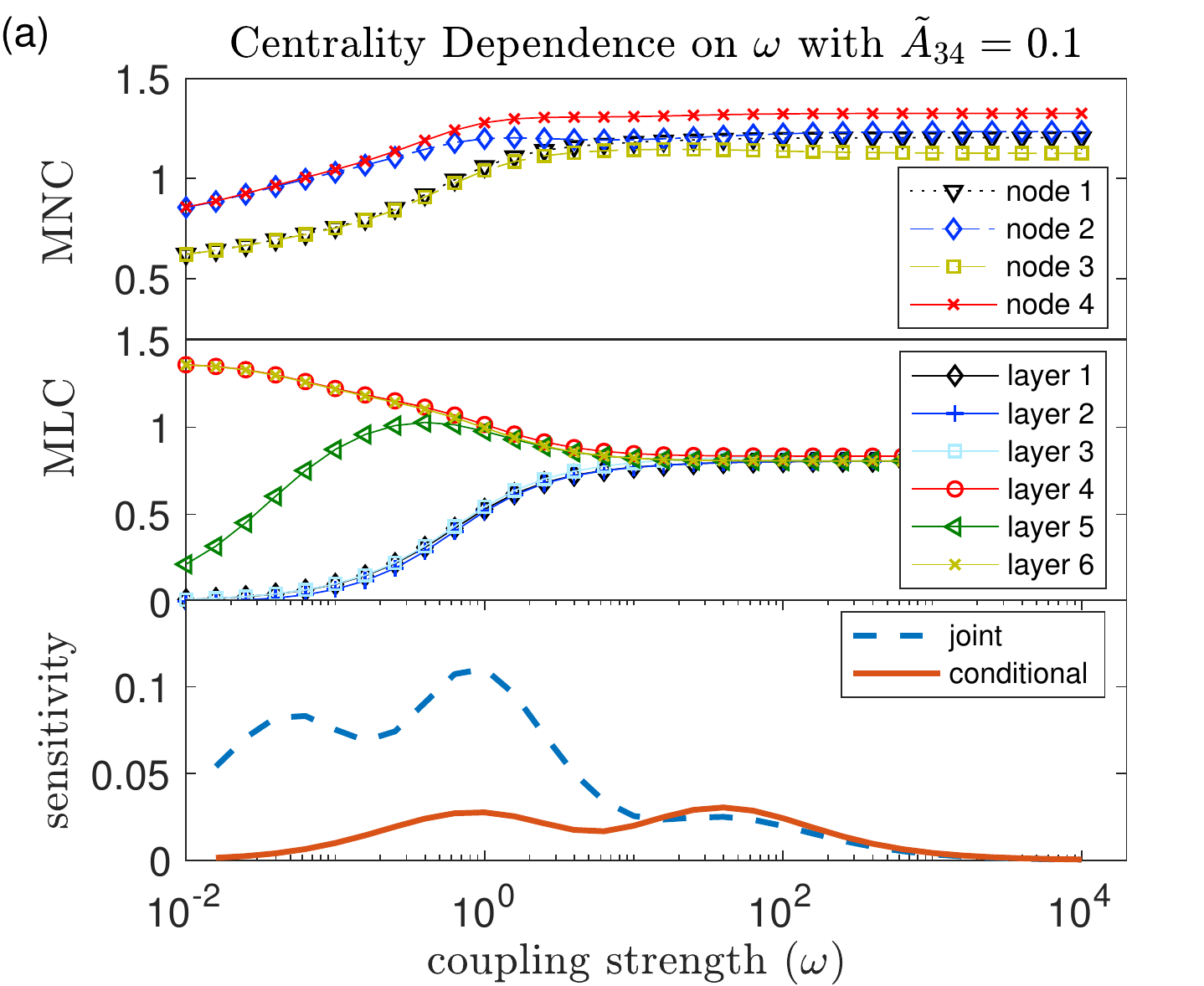}
\includegraphics[width=.45\linewidth]{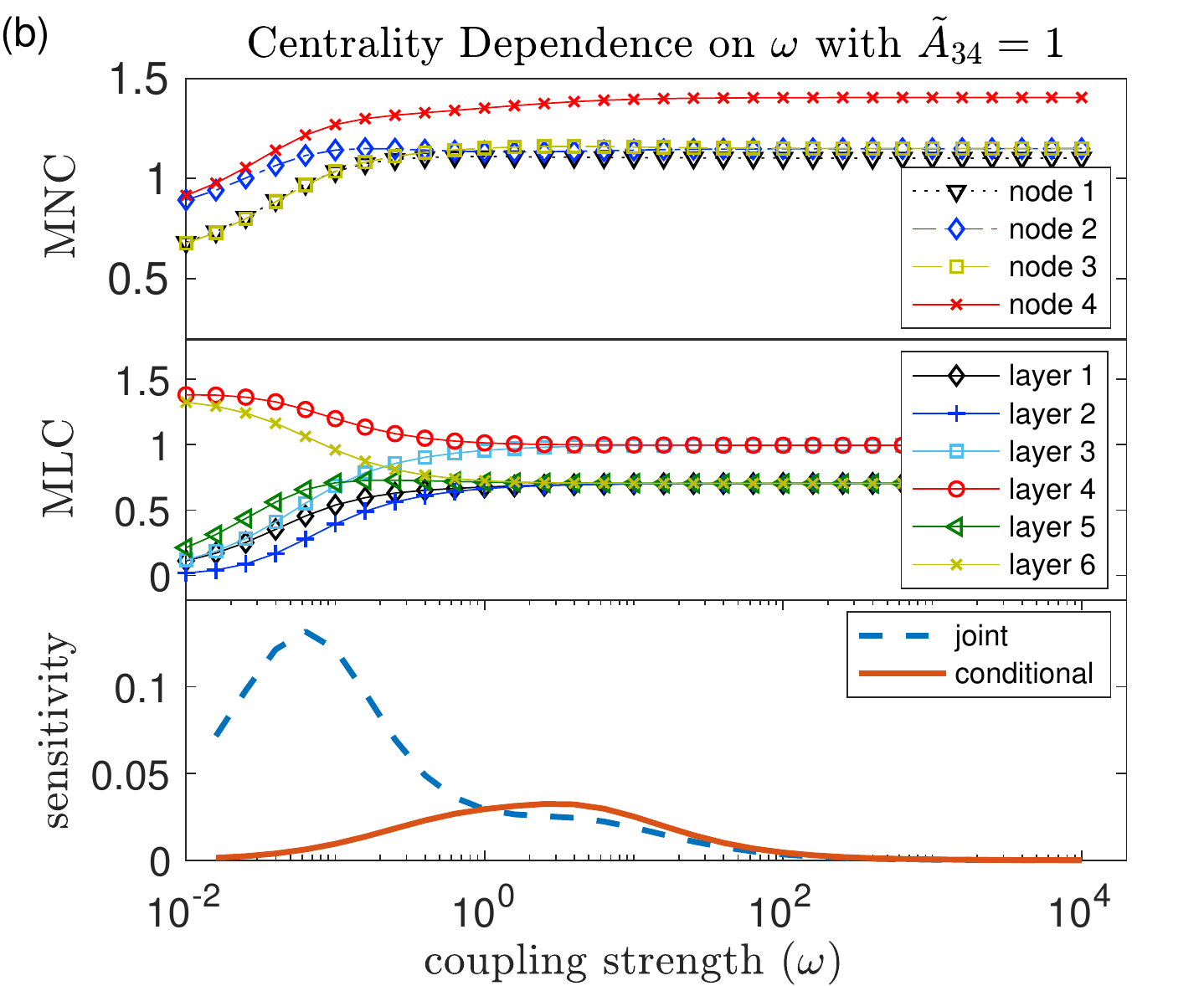}
\caption{Eigenvector supracentrality results for our pedagogical multiplex network in which the coupling strength between layers $3$ and $4$ differs from 
that between the other layers. In panels (a) and (b), we report the same results as in Fig.~3(b) {of} the main text (for which we used $\tilde{A}_{34}=\tilde{A}_{43} = 0.01$), except that now we use 
(a) $\tilde{A}_{34}=\tilde{A}_{43} = 0.1$ and 
(b) $\tilde{A}_{34}=\tilde{A}_{43} = 1$. In panel (b), note that the curves for ``sensitivity'', which we measure in terms {of the} stepwise magnitudes of change (specifically, $\|{\bf W}(\omega_s) - {\bf W}(\omega_{s-1}) \|_F$ and $\|{\bf Z}(\omega_s) - {\bf Z}(\omega_{s-1}) \|_F$), are no longer bimodal. 
}
\label{fig:sweep_SI}
\end{figure}

In Fig.~\ref{fig:sweep_SI}, we now explore two other choices of interlayer-coupling strengths between layers 3 and 4: (a) $\tilde{A}_{34}=\tilde{A}_{43} = 0.1$ and (b) $\tilde{A}_{34}=\tilde{A}_{43} = 1$. We plot the marginal node centralities (MNCs), marginal layer centralities (MLCs), and ``sensitivity'' of the joint and conditional centralities across a range of coupling strengths $\omega$. Our measures of sensitivity are the stepwise magnitudes of change (i.e., $\|{\bf W}(\omega_s) - {\bf W}(\omega_{s-1}) \|_F$ and $\|{\bf Z}(\omega_s) - {\bf Z}(\omega_{s-1}) \|_F$). In panel (b), note that the curves are no longer bimodal, implying that a stable intermediate regime vanishes as we increase $\tilde{A}_{34}$. Intuitively, this occurs because the network that is associated with the interlayer-adjacency matrix $\tilde{{\bf A}}$ of Fig.~1(a) no longer has two well-separated communities if $\tilde{A}_{34}=\tilde{A}_{43} = 1$.


\section{Extended Study of European Airport Rankings}\label{sec:euro}
%

We now discuss additional results for our study of a European airline transportation multiplex network (with data from \cite{cardillo2013emergence}) from Sec.~5.1 of the main text. Recall that this network includes $N=417$ European airports, which are
in the largest strongly connected component of an aggregation (which we obtain from summing the layers' adjacency matrices) of the multiplex network. There are $T=37$ layers, each of which encodes the flight patterns between airports for a single airline.

\begin{table}[b!]
\caption{European airports with the top marginal node centralities (MNCs) for coupling strengths $\omega\in\{0.01,1,100\}$, which are in
the regimes of weak ($\omega = 0.01$), intermediate ($\omega = 1$), and strong ($\omega = 100$) coupling. 
We show results for when the layers' centrality matrices are PageRank matrices and the interlayer-adjacency matrix is $\tilde{\bm{A}}=\bm{1}\bm{1}^T$. We indicate airports using their International Civil Aviation Organization (ICAO) codes. 
}
\centering{
\small{
\begin{tabular}{cccc} 
\begin{tabular}{c}
~\\
\hline\hline 
Rank \\
[0.5ex] 
\hline  
1 \\
2 \\
3 \\
4 \\
5 \\
6 \\
7 \\
8 \\
9 \\
10 \\
[1ex]  
\hline\hline
\end{tabular}&
\begin{tabular}{c}
$\omega=10^{-2}$\\
\begin{tabular}{cc}
\hline\hline 
Airport  & MNC\\  [0.5ex] 
\hline  
EHAM& 0.406\\
LOWW& 0.373\\
LTBA& 0.372\\
EGKK& 0.365\\
LEMD& 0.363\\
LTFJ& 0.344\\
LFPG& 0.337\\
LGAV &0.333\\
EGLL &0.328\\
EIDW &0.328\\
[1ex]  
\hline \hline
\end{tabular}
\end{tabular}&
\begin{tabular}{c}
$\omega=10^{0}$\\
\begin{tabular}{cc}
\hline\hline 
Airport & MNC  \\  [0.5ex] 
\hline  
LTBA &0.802\\
EBLG& 0.732\\
LTFJ &0.700\\
EVRA &0.695\\
EHAM &0.662\\
EGKK &0.653\\
LOWW &0.633\\
EIDW& 0.586\\
EGSS &0.583\\
LEMD &0.554\\
[1ex]  
\hline \hline
\end{tabular}
\end{tabular}&
\begin{tabular}{c}
$\omega=10^2$\\
\begin{tabular}{cc}
\hline\hline 
Airport  & MNC  \\  [0.5ex] 
\hline  
LTBA &0.929\\
EBLG &0.866\\
EVRA &0.818\\
LTFJ &0.791\\
EHAM &0.725\\
LOWW& 0.699\\
EGKK& 0.698\\
EIDW &0.656\\
EGSS &0.631\\
LEMD &0.596\\
[1ex]  
\hline \hline
\end{tabular}
\end{tabular}
\end{tabular}}
\label{table:tops2} }
\end{table}

In Sec.~5.1 of main text, we studied supracentralities for the European airline transportation multiplex network using the layers' adjacency matrices as their centrality matrices $ {\bf C}^{(t)} =  {\bf A}^{(t)} $ (so these are eigenvector supracentralities), and we found that the Ryanair network dominates the supracentralities in the weak-coupling limit because of eigenvector localization. Specifically, as we illustrated in Fig.~4(a) of the main text, $\mathbbm{v}(\omega)$ localizes onto layer $t=2$. \new{Corollary 4.8} of the main text describes this phenomenon, which occurs because the centrality matrix for Ryanair has the largest spectral radius.

We now study supracentrailities when the layers' centrality matrices are PageRank matrices. (See \new{Definition 2.6} {of} the main text.) \new{We add a self-edge to each node to ensure that there are no dangling nodes.} In Table~\ref{table:tops2}, we list the airports with the largest MLCs for small, intermediate, and large values of the coupling strength $\omega$.  Unsurprisingly, there is some overlap with the top-ranked airports in Table 2 of the main text (which is based on eigenvector supracentralities). For example, LEMD (Adolfo Su\'arez Madrid--Barajas Airport) makes the top-10 list for all values of $\omega$ in both tables. However, most of the top-ranked airports are different. In particular, for $\omega\ge 0.1$, the top-ranked airports are LTBA, EBLG, LTFJ, and EVRA for PageRank supracentrality; none of these airports appear in Table 2 of the main text. It is unsurprising that we observe these differences, as it is well-known for monolayer networks that PageRank and eigenvector centrality identify different types of node importances (although they are correlated with each other).

\begin{figure}[t]
\centering
\includegraphics[width=.95\linewidth]{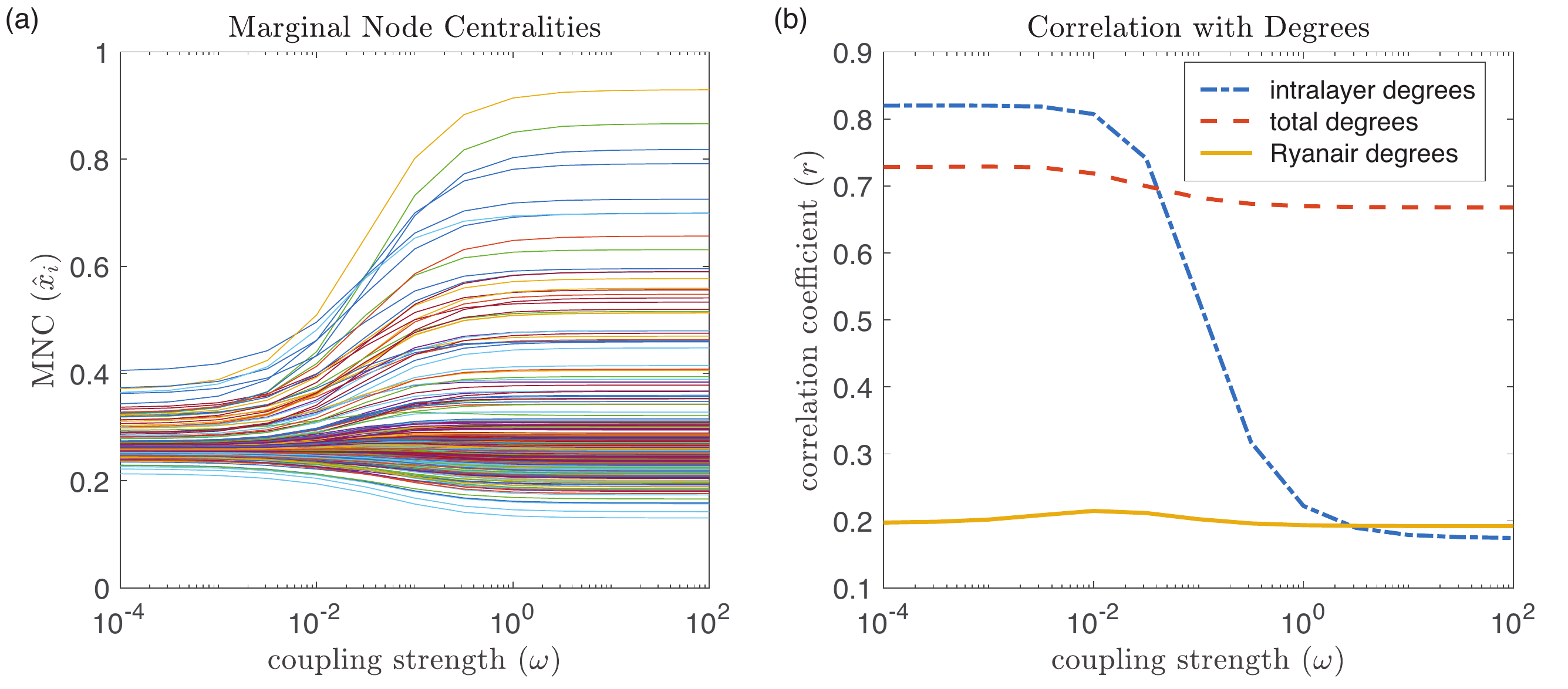}
\vspace{-0.3cm}
\caption{Supracentralities for the multiplex European airline network when the layers' centrality matrices are given by PageRank matrices.
(a)~Airport MLCs versus intralayer coupling strength $\omega$.
(b)~Pearson correlation coefficients for comparing PageRank supracentralities to three different notions of node degree. (See the description in the caption of Fig.~6(b) of the main text.)
}\label{fig:mux_deg_eig}
\end{figure}

We can gain insight into one of the main causes of these differences by examining the limit of small $\omega$. In Fig.~6(b) of the main text, we observed that the conditional centralities correlate very strongly with the intralayer node degrees in the Ryanair layer (but not in the other layers). However, as we can see in Fig.~\ref{fig:mux_deg_eig}(b), the PageRank MNCs paint a very different picture. 
Specifically, for small values of $\omega$, the conditional node centralities correlate strongly with the intralayer node degrees in all layers. 
In other words, for eigenvector supracentrality, the Ryanair layer dominates, and the supracentralities largely reflect the intralayer node degrees of this single layer. By contrast, for PageRank supracentralities, the Ryanair layer is not dominant. Whether 
there is such domination or not depends on whether
eigenvector localization occurs as $\omega\to0^+$, as described by \new{Thm. 4.4 and Corollary 4.8} of the main text. In this example, localization occurs for eigenvector supracentrality but not for PageRank supracentrality. There is no such localization in the latter case, because PageRank matrices are transition matrices for Markov chains and thus \new{they all} have a spectral radius of $1$.



As a final experiment with the multiplex airline network, we compare our results to calculations that use \emph{PageRank versatility} \cite{de2015ranking}, a different extension of PageRank for multiplex networks. We implement PageRank versatility by building a supra-adjacency matrix $\mathbb{A} = \text{diag}[{\bf A}^{(1)},\dots,{\bf A}^{(t)}] + \omega \tilde{\bm{A}} \otimes {\bf I}$ with interlayer edges of weight $\omega$. Interpreting $\mathbb{A}$ as an ordinary adjacency matrix (i.e., ignoring the fact that some edges are intralayer edges and others are interlayer ones, we construct its associated PageRank matrix by substituting $\mathbb{A}\mapsto {\bf A}$ in \new{Definition 2.6} of the main text.
The dominant eigenvector of the resulting matrix gives centralities for node-layer pairs. It is similar to our notion of joint centrality (see \new{Definition 3.3}), and both yield centralities for node-layer pairs. 
 We calculate the PageRank versatility of each node by summing the eigenvector entries that are associated with the node-layer pairs for that node. 
 %
 PageRank versatility is similar to our notion of MNCs
 (see \new{Definition 3.5} of the main text), and one calculates each of them by summing the centralities of node-layer pairs for each node.
 %
 In Table~\ref{table:tops3}, we list the top-ranked airports according to PageRank versatility. The list includes many of the same top-ranked airports that we identified using eigenvector supracentrality (see Table 3 of the main text) and PageRank supracentrality (see Table \ref{table:tops2}). 


\begin{table}[h!]
\caption{
{European airports with the top PageRank versatilities \cite{de2015ranking} for supracentrality matrices that we construct using interlayer-coupling strengths $\omega\in\{0.01,1,100\}$. We use a teleportation parameter of $\sigma=0.85$.
}}
\centering{{
\footnotesize{
\begin{tabular}{cccc} 
\begin{tabular}{c}
~\\
\hline\hline 
Rank \\
[0.5ex] 
\hline  
1 \\
2 \\
3 \\
4 \\
5 \\
6 \\
7 \\
8 \\
9 \\
10 \\
[1ex]  
\hline\hline
\end{tabular}&
\begin{tabular}{c}
$\omega=10^{-2}$\\
\begin{tabular}{cc}
\hline\hline 
Airport  & MNC\\  [0.5ex] 
\hline  
EHAM& 0.429\\
LEMD& 0.386\\
LOWW& 0.384\\
LTBA& 0.374\\
EGKK& 0.373\\
EDDM& 0.359\\
LGAV& 0.353\\
LFPG &0.349\\
EDDF &0.343\\
EGGS &0.342\\
[1ex]  
\hline \hline
\end{tabular}
\end{tabular}&
\begin{tabular}{c}
$\omega=10^{0}$\\
\begin{tabular}{cc}
\hline\hline 
Airport & MNC  \\  [0.5ex] 
\hline  
EGSS &0.307\\
EHAM& 0.306\\
EDDM &0.306\\
EGKK &0.306\\
LTBA &0.305\\
EDDF &0.305\\
LEMD &0.305\\
EIDW& 0.304\\
LOWW &0.304\\
LFPG &0.302\\
[1ex]  
\hline \hline
\end{tabular}
\end{tabular}&
\begin{tabular}{c}
$\omega=10^2$\\
\begin{tabular}{cc}
\hline\hline 
Airport  & MNC  \\  [0.5ex] 
\hline  
EGSS &0.298\\
LTBA &0.298\\
EDDM &0.298\\
EDDF &0.298\\
EGKK  &0.298\\
EHAM& 0.298\\
LOWW& 0.298\\
EIDW &0.298\\
LFPG &0.298\\
 LEMD&0.298\\
[1ex]  
\hline \hline
\end{tabular}
\end{tabular}
\end{tabular}}}
\label{table:tops3} }
\end{table}


\section{Extended Study of   U.S. Mathematical-Sciences Doctoral Program Rankings}\label{sec:phd}
%
We now present additional results for our study (see Sec.~5.2 of the main text) of supracentralities in a temporal network of the graduation and hiring of mathematical-science Ph.D. recipients between U.S. universities. (The data set, which was released with \cite{taylor2017eigenvector} and is available at \cite{MGP_data}, was compiled from the Mathematics Genealogy Project \cite{mgp}.) In Table~\ref{table:MGP1}, we list the universities with the top MNCs when the layers' centrality matrices are given by their authority matrices. (See Definition~2.4 of the main text.) The interlayer-adjacency matrix $\tilde{\bm{A}}$ is given by Eq.~(5.5) of the main text. We show results for several choices of the layer teleportation parameter $\gamma$ and interlayer-coupling strength $\omega$. 
Although the ordering of the top-ranked schools depends sensitively on the values of $\gamma$ and $\omega$, we typically obtain the same set of universities near the top. MIT, for example, is almost always ranked first throughout $(\gamma,\omega)$ parameter space.

\begin{table}[h]
\caption{
Top MNCs (see \new{Definition 3.5} {of} the main text) for U.S. doctoral programs in the mathematical sciences when the layers' centrality matrices are authority matrices and the interlayer-adjacency matrix is given by Eq.~(5.5) of the main text. We show results for three choices of $(\gamma,\omega)$.
}
{\centering 
\scriptsize{
\begin{tabular}{c c cc} 
~&   $(\gamma,\omega)=(10^{-2},1)$ &  $(\gamma,\omega)=(10^{-2},10^{2})$ &   $(\gamma,\omega)=(10^{-3},10^{2})$ \\
\begin{tabular}{c } 
\hline\hline 
Rank  \\ [0.5ex] 
\hline 
1  \\ 
2  \\
3  \\
4   \\
5  \\ 
6  \\ 
7   \\ 
8   \\ 
9   \\ 
10   \\ 
[1ex] 
\hline 
\end{tabular}
&
\begin{tabular}{c c } 
\hline\hline 
University & $\hat{x}_i$   \\ [0.5ex] 
\hline 
MIT & 0.91    \\ 
U Washington & 0.23  \\
Boston U & 0.15 \\
U Michigan & 0.12   \\
Brown & 0.12    \\ 
UCLA & 0.111   \\ 
Carnegie Mellon & 0.11    \\ 
Purdue & 0.11    \\ 
USC & 0.11   \\ 
U of Georgia & 0.11    \\ 
[1ex] 
\hline 
\end{tabular}
&
\begin{tabular}{c c } 
\hline\hline 
University & $\hat{x}_i$   \\ [0.5ex] 
\hline 
MIT & 5.28    \\ 
UC Berkeley  & 2.28  \\
 Stanford & 1.84 \\
Princeton  & 1.42   \\
 Harvard& 1.28    \\ 
Cornell & 1.23  \\ 
UIUC  & 1.18    \\ 
Washington & 1.13    \\ 
U Michigan & 1.12   \\ 
UCLA & 1.09    \\ 
[1ex] 
\hline 
\end{tabular}
&
\begin{tabular}{c c } 
\hline\hline 
University & $\hat{x}_i$   \\ [0.5ex] 
\hline 
MIT & 3.47    \\ 
UC Berkeley  &1.72  \\
Stanford &1.28 \\
Harvard & 0.97   \\
Princeton & 0.96    \\ 
Cornell & 0.89   \\ 
UIUC & 0.77    \\ 
UCLA  & 0.75    \\ 
 Wisconsin-Madison & 0.74   \\ 
U Michigan & 0.66    \\ 
[1ex] 
\hline 
\end{tabular}
\end{tabular}
}}
\label{table:MGP1} 
\end{table}

\begin{figure}[h!]
\centering
\includegraphics[width=.95\linewidth]{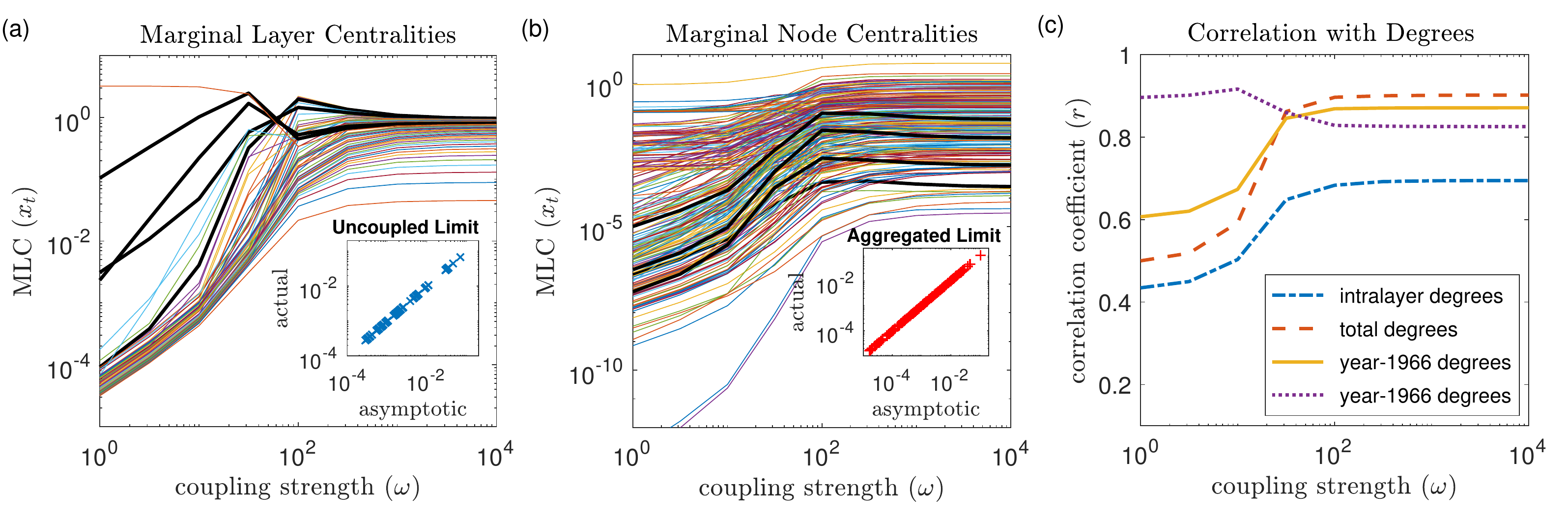}
\vspace{-.3cm}
\caption{
Supracentralities of the Ph.D. exchange network with the layers' centrality matrices given by their authority matrices and the interlayer-adjacency matrix given by Eq.~(5.5) of the main text with layer teleportation parameter $\gamma = 10^{-2}$. 
(a) MLCs versus $\omega$. (b) MNCs versus $\omega$.
The insets in panels (a) and (b) compare the calculated
conditional node centralities for $\omega=1 $ and $\omega=10^{4}$, respectively, to the asymptotic values from \new{Thms.~4.4 and 4.13}, respectively.
(c) We compute Pearson correlation coefficients to measure the similarity between various authority supracentralities and four different notions of multiplex node degree: {({dot-dashed blue} curve) intralayer degrees $d_i^{(t)} = \sum_j A_{ij}^{(t)}$ versus conditional node centralities $Z_{it} $; 
({dashed red} curve) total degrees $\overline{d}_i   =\sum_t d_i^{(t)} $ versus the sum $\sum_t Z_{it} (\omega)$ of the conditional node centralities;   
({solid} gold curve) degrees $d_i^{(2)} = \sum_j A_{ij}^{(2)}$ in the Ryanair layer versus  $\sum_t Z_{it} (\omega)$; and 
(dotted purple curve)  the values $\sum_j {C}_{ij}^{(1966)} = \sum_{k,j} {A}_{ki}^{(1966)}{A}_{kj}^{(1966)}$  versus  $\sum_t Z_{it} (\omega)$.}
}
\label{fig:causal_MGP}
\end{figure}

In Fig.~\ref{fig:causal_MGP}, we illustrate the effect of $\omega$ on the authority supracentralities for a layer teleportation-parameter value of $\gamma=10^{-2}$. In panels (a) and (b), we show the layers' MLCs and nodes' MNCs, respectively. We observe three qualitative regimes: weak, intermediate, and strong coupling. In the insets of Fig.~\ref{fig:causal_MGP}(a) and Fig.~\ref{fig:causal_MGP}(b), we compare calculated conditional node centralities for $\omega=1$ and $\omega=10^{4}$, respectively, to the asymptotic values from \new{Thms.~4.4 and 4.13}, respectively. 

In Fig.~\ref{fig:causal_MGP}(c), we plot (as a function of $\omega$) the Pearson correlation coefficient $r$ between authority supracentralities and the same three notions of node degree as for Fig.~6(b) {of} the main text and Fig.~\ref{fig:mux_deg_eig}(b).
%
Additionally, the dotted purple curve is the Pearson correlation coefficient between the MNCs and the values $\sum_j {C}_{ij}^{(1966)} = \sum_{k,j} {A}_{ki}^{(1966)}{A}_{kj}^{(1966)}$, which give a first-order approximation to the authority scores for layer $t=1966$. 
 (See footnote 3 of the main text.) 
 For very large values of $\omega$, 
 the authority supracentralities correlate most strongly with the nodes' total degrees ({dashed red curve}).  For small $\omega$, the authority supracentralities correlate most strongly with the row sums of the matrices $\bm{A}^{(1966)}$ (solid gold curve) and 
$\bm{C}^{(1966)}$ for layer $t=1966$ ({dotted purple} curve). This occurs because the spectral radii of these matrices are larger than those of the matrices that are associated with the other layers, and \new{there is} localization \new{onto} layer $t=1966$ (the dominating layer), as characterized by \new{Corollary 4.8} of the main text. 




